\documentclass{aa}

\usepackage{natbib}
\usepackage{graphicx}
\usepackage{txfonts}
\usepackage{hyperref}
\makeatletter
\renewcommand*\aa@pageof{, page \thepage{} of \pageref*{LastPage}}
\makeatother

\newcommand{\hrieuv}{HRI\textsubscript{EUV}\xspace}
\newcommand{\hrilya}{HRI\textsubscript{Lya}\xspace}

\usepackage[normalem]{ulem}

\begin{document}

   \title{First Perihelion of EUI on the Solar Orbiter mission}
   \author{
        D. Berghmans\inst{\ref{i:rob}}\fnmsep\thanks{Corresponding author: David Berghmans \email{david.berghmans@oma.be}}
        \and
        P. Antolin\inst{\ref{i:unn}}
        \and
        F. Auch\`ere\inst{\ref{i:ias}}
        \and
        R. Aznar Cuadrado\inst{\ref{i:mps}}
        \and
        K. Barczynski\inst{\ref{i:pmod},\ref{i:eth}}
        \and
        L. P. Chitta\inst{\ref{i:mps}}
        \and
        S. Gissot\inst{\ref{i:rob}}
       \and
        L. Harra\inst{\ref{i:pmod},\ref{i:eth}}
        \and
        Z. Huang\inst{\ref{i:mps}}
        \and
        M. Janvier\inst{\ref{i:esa}, \ref{i:ias}}
        \and
        E. Kraaikamp\inst{\ref{i:rob}}
        \and
        D.~M. Long\inst{\ref{i:mssl}}
        \and
        S. Mandal\inst{\ref{i:mps}}
       \and
        M. Mierla\inst{\ref{i:rob}}
        \and
         S. Parenti\inst{\ref{i:ias}, \ref{i:rob}}
        \and
        H. Peter\inst{\ref{i:mps}}
        \and
         L. Rodriguez\inst{\ref{i:rob}}
        \and
        U. Sch\"uhle\inst{\ref{i:mps}}
        \and
        P.~J. Smith\inst{\ref{i:mssl}}
        \and
        S.~K.~Solanki\inst{\ref{i:mps}}
        \and
        K. Stegen\inst{\ref{i:rob}}
        \and
        L. Teriaca\inst{\ref{i:mps}}
        \and
        C. Verbeeck\inst{\ref{i:rob}}
        \and
        M.~J. West\inst{\ref{i:swri}}
        \and
        A.~N.~Zhukov\inst{\ref{i:rob},\ref{i:sinp}}
        \and
        T. Appourchaux\inst{\ref{i:ias}}
        \and
        G. Aulanier\inst{\ref{i:ObservatoireDeParis},\ref{i:RoCS}}
        \and
        E. Buchlin\inst{\ref{i:ias}}
        \and
        F. Delmotte\inst{\ref{i:iogs}}
        \and
        J.~M. Gilles\inst{\ref{i:csl}}
        \and
        M. Haberreiter\inst{\ref{i:pmod}}
        \and
        J.-P. Halain\inst{\ref{i:csl},\ref{i:esa}}
        \and
        K. Heerlein\inst{\ref{i:mps}}
        \and
        J.-F. Hochedez\inst{\ref{i:aester},\ref{i:latmos}}
        \and
        M. Gyo\inst{\ref{i:pmod}}
        \and
        S. Poedts\inst{\ref{i:kuleuven},\ref{i:umcs}}
        \and
        P. Rochus\inst{\ref{i:csl}}
       }
        \institute{
            Solar-Terrestrial Centre of Excellence -- SIDC, Royal Observatory of Belgium, Ringlaan -3- Av. Circulaire, 1180 Brussels, Belgium\label{i:rob}
            \and
            Department of Mathematics, Physics and Electrical Engineering, Northumbria University, Newcastle upon Tyne, NE1 8ST, UK\label{i:unn}
            \and
            Institut d'Astrophysique Spatiale, CNRS, Univ. Paris-Sud, Universit\'e Paris-Saclay, B\^at.\ 121, 91405 Orsay, France\label{i:ias}
            \and
             Max Planck Institute for Solar System Research, Justus-von-Liebig-Weg 3, 37077 G\"ottingen, Germany\label{i:mps}
            \and
            Physikalisch-Meteorologisches Observatorium Davos, World Radiation Center, 7260, Davos Dorf, Switzerland\label{i:pmod}
            \and
            ETH Z\"urich, Institute for Particle Physics and Astrophysics , Wolfgang-Pauli-Strasse 27, 8093 Z\"urich \label{i:eth}
            \and
            European Space Agency, ESTEC, Keplerlaan 1, PO Box 299, NL-2200 AG Noordwijk, The Netherlands\label{i:esa}
            \and
            UCL-Mullard Space Science Laboratory, Holmbury St.\ Mary, Dorking, Surrey, RH5 6NT, UK\label{i:mssl}
            \and
            Southwest Research Institute, 1050 Walnut Street, Suite 300, Boulder, CO 80302, USA\label{i:swri}
            \and
            Skobeltsyn Institute of Nuclear Physics, Moscow State University, 119992 Moscow, Russia\label{i:sinp}
            \and
            Sorbonne Université, Observatoire de Paris - PSL, École Polytechnique, Institut Polytechnique de Paris, CNRS, Laboratoire de physique des plasmas (LPP), 4 place Jussieu, F-75005 Paris, France\label{i:ObservatoireDeParis}
            \and
            Rosseland Centre for Solar Physics, University of Oslo, P.O. Box 1029, Blindern, NO-0315 Oslo, Norway\label{i:RoCS}
            \and
            Laboratoire Charles Fabry, Institut d'Optique Graduate School, Universit\'e Paris-Saclay, 91127 Palaiseau Cedex, France\label{i:iogs}
            \and
            Centre Spatial de Li\`ege, Universit\'e de Li\`ege, Av. du Pr\'e-Aily B29, 4031 Angleur, Belgium\label{i:csl}
            \and
            %Centre for Mathematical Sciences, University of Cambridge, Wilberforce Road, Cambridge CB3 0WA, UK\label{i:damtp}
            %\and
            %Centre National d'Etudes Spatiales, 18, avenue Edouard Belin, %31401 Toulouse Cedex 9, France\label{i:cnes}
            %\and
            AESTER INCOGNITO, 75008 Paris, France\label{i:aester}
            \and
            LATMOS, CNRS - UVSQ - Sorbonne Université, 78280, Guyancourt, France\label{i:latmos}
            \and
            Centre for mathematical Plasma Astrophysics, KU Leuven, 3001 Leuven, Belgium\label{i:kuleuven}
            \and
            Institute of Physics, University of Maria Curie-Sk{\l}odowska, Pl.\ M.\ Curie-Sk{\l}odowskiej 5, 20-031 Lublin, Poland\label{i:umcs}
             }

   \date{Received ; accepted }

% 5 {} token are mandatory

  \abstract
   {The Extreme Ultraviolet Imager (EUI), onboard Solar Orbiter consists of three telescopes: the two High Resolution Imagers in EUV (\hrieuv) and in Lyman-$\alpha$ (\hrilya), and the Full Sun Imager (FSI). Solar Orbiter/EUI started its Nominal Mission Phase on 2021 November 27.}
   {EUI images from the largest scales in the extended corona off limb, down to the smallest features at the base of the corona and chromosphere. EUI is therefore a key instrument for the connection science that is at the heart of the Solar Orbiter mission science goals.}
   {The highest resolution on the Sun is achieved when Solar Orbiter passes through the perihelion part of its orbit. On 2022 March 26, Solar Orbiter reached for the first time a distance to the Sun close to 0.3\,au. No other coronal EUV imager has been this close to the Sun.}
   {We review the EUI data sets obtained during the period 2022 March-April, when Solar Orbiter quickly moved from alignment with the Earth (2022 March 6), to perihelion (2022 March 26), to quadrature with the Earth (2022 March 29). We  highlight the first observational results in these unique data sets and we report on the in-flight instrument performance.}
   {EUI has obtained the highest resolution images ever of the solar corona in the quiet Sun and polar coronal holes. Several active regions were imaged at unprecedented cadences and sequence durations. We identify in this paper a broad range of features that require deeper studies. Both FSI and \hrieuv operate at design specifications but \hrilya suffered from performance issues near perihelion. We conclude emphasising the EUI open data policy and encouraging further detailed analysis of the events highlighted in this paper.}
   \keywords{Sun: UV radiation -- Sun: transition region -- Sun: corona --  Instrumentation: high angular resolution}
\maketitle

\section{Introduction\label{sec:introduction}}

The launch of the Atmospheric Imaging Assembly (AIA) onboard the Solar Dynamics Observatory (SDO, \cite{2012SoPh..275....3P}) in 2010 heralded the era of continuous full disc coronal imaging at high spatial resolution.
In normal mode, AIA images are produced with a cadence of 12\,s
at a spatial resolution of $1.5\arcsec$, over a field of view (FOV) of $(41\arcmin)^2$
\citep{2012SoPh..275...17L}.
% Lemen et al, 2012)
Recent developments in coronal imagers have included increased fields of view and higher spatial resolution. SWAP on PROBA2
\citep{2013SoPh..286...43S, 2013SoPh..286...67H}
% Seaton 2013, Halain 2013)
and SUVI on GOES
\citep{2022SpWea..2003044D}
%Darnel et al 2022)
have boosted observations with FOVs of $(54\arcmin)^2$ and $(53\arcmin)^2$ respectively. Thanks to these larger FOVs, both SWAP and SUVI image the EUV structures and dynamics well beyond the AIA FOV, into what has become known as the Middle Corona
\citep{2021NatAs...5.1029S,2022arXiv220804485W,2022Chitta..NatA}.
%(Seaton 2021, Matt 2022).
Meanwhile, the sounding rocket Hi-C
\citep{2014SoPh..289.4393K}
%(Kobayashi et al., 2014)
pushed the limits in terms of spatial resolution. In its second successful flight
\citep{2019SoPh..294..174R},
%(Laurel 2019)
Hi-C took (subfield) images of an active region at a cadence of 4\,s and a spatial resolution better than $0.46\arcsec$ (330\,km on the Sun).

Solar Orbiter \citep{Muller2020} is in a highly elliptical orbit with perihelia below 0.3\,au and, in later years of the nominal mission, well out of the ecliptic, beyond $30^\circ$ solar latitude.
The EUI instrument \citep{Rochus2020} onboard Solar Orbiter will use this unique orbit to observe the Sun from different vantage points through three separate telescopes, imaging the outer solar atmosphere at an even higher spatial resolution than Hi-C, and over wider fields of view than SUVI and SWAP, further extending the middle corona discovery space.

The first EUI telescope, the Full Sun Imager (FSI) is a one-mirror telescope taking alternating images in the 17.4\,nm  and 30.4\,nm passbands.
For a coronal EUV imager, FSI has an unprecedented large FOV:  $(228\arcmin)^2$, which has a significant overlap \citep{Auchere2020} with the Solar Orbiter coronagraph Metis \citep{Antonucci2020}. At perihelion, this FOV corresponds to $(4~R_\odot)^2$ such that the full solar disc is always seen, even at maximal offpoint (1~$R_\odot$). This FOV is significantly wider than the $(3.34~R_\odot)^2$ of EUVI \citep{Howard2008} or the $(3.38~R_\odot)^2$ of SWAP \citep{Seaton2013}.
When at 1\,au (near aphelion), this FOV corresponds to (14.3~$R_\odot$)$^2$ providing unique opportunities to image the Middle Corona and eruptions that transit through this region.

The other EUI telescopes are the two High Resolution Imagers (HRIs), \hrieuv and \hrilya, which are two-mirror optical systems imaging through EUV and  hydrogen Lyman-$\alpha$ passbands respectively.
\hrieuv images the corona at 17.4\,nm, which corresponds to the 17.4\,nm channel of FSI. \hrilya, which images in the Lyman-$\alpha$ line, shares its resonance formation process for hydrogen with the 30.4\,nm channel of FSI for helium.
The \hrieuv plate scale is 0.492\arcsec, the \hrilya plate scale is 0.514\arcsec. At the 2022 March 26 perihelion, Solar Orbiter reached a distance of  0.323\,au from the Sun, giving (single) pixel values on the Sun of (115\,km)$^2$ for \hrieuv, and (120\,km)$^2$ for \hrilya. The actual spatial resolution of the telescopes is discussed in Section \ref{section:Performance}.
Both HRI cameras are capable of operating at cadences in the 1\,s range, over 2048$\times$2048 pixel arrays. The HRIs image through a $17'\times17'$ FOV, corresponding to (1 $R_\odot$)$^2$, when observing at 1\,au, and (0.28 $R_\odot$)$^2$ at perihelion.

Following its launch on 2020 February 10, Solar Orbiter spent 4 months in the Near Earth Commissioning Phase, followed by 17 months of Cruise Phase. During the Cruise Phase only the in-situ instruments were collecting science grade data, while the remote sensing instruments were undergoing extended testing in preparation for the science phase of the mission. The Nominal Mission Phase of Solar Orbiter started on 2021 November 27. During the Nominal Mission Phase, the remote sensing instruments run a non-stop synoptic observation program interleaved three times per orbit with 10 days periods of enhanced observational activity. These periods are called ``Remote Sensing Windows'' (RSWs) and are typically scheduled at the perihelion of the orbit of Solar Orbiter and at the times of minimum and maximum latitude.

\begin{figure}[ht]
\centering
\includegraphics[width=0.48\textwidth]{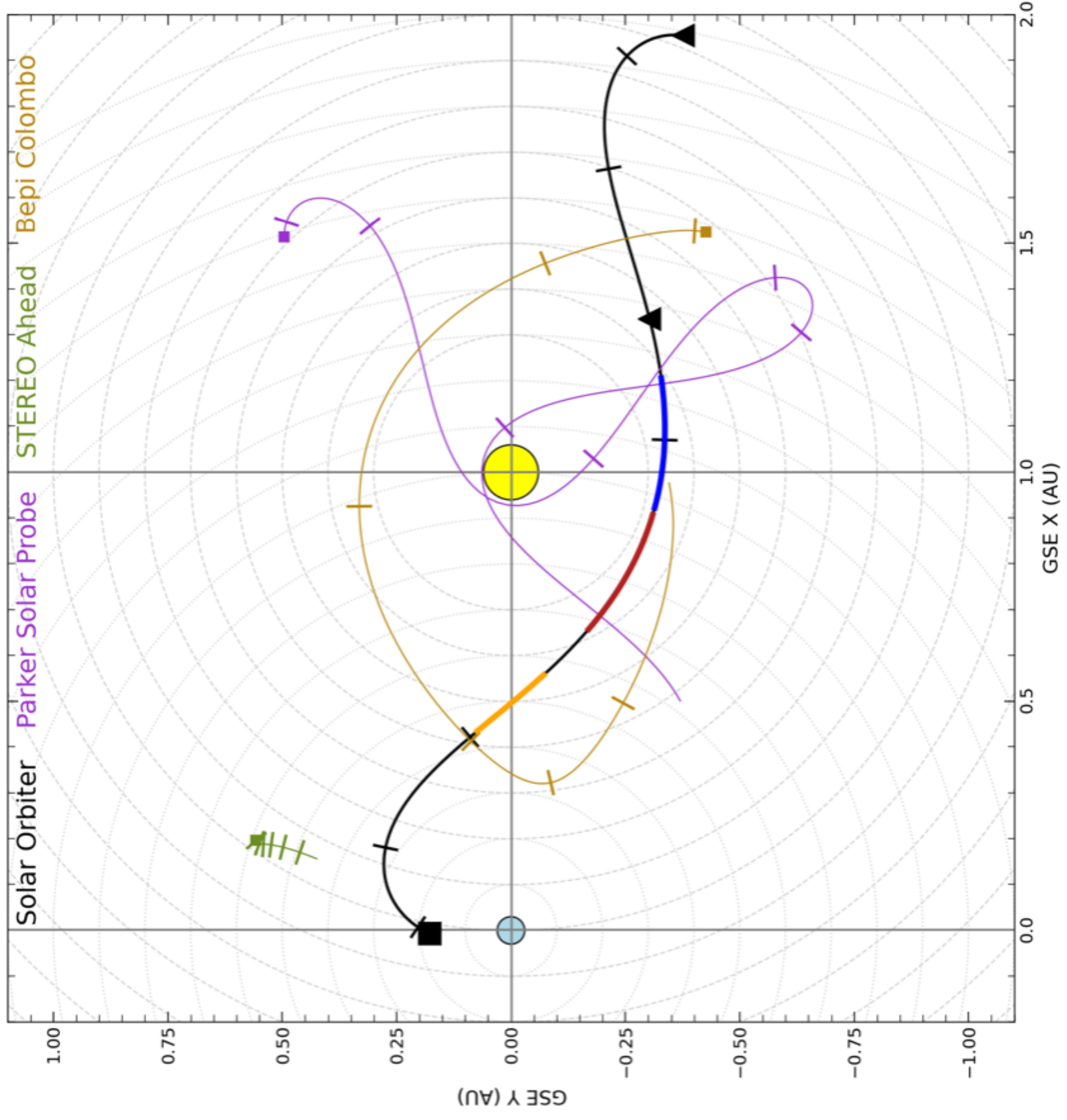}
\caption{Trajectory of Solar Orbiter in Geocentric Solar Ecliptic (GSE) coordinates in black, starting at 2021 December 27 (square symbol). The Remote Sensing Windows (RSW) correspond to the orange, red, and blue parts of the trajectory. In this paper we cover the period from 2022 March 02 (beginning of RSW1, orange) till 2022 April 06 (end of RSW3, blue). The perihelion occurred at the end of the RSW2 (red), around 2022 March 26. The trajectories of the ESA mission Bepi Colombo and the NASA missions STEREO-A, and Parker Solar Probe are indicated brown, green, and purple respectively.  \label{fig:LTP6orbit.pdf}}
\end{figure}

In this paper we present an overview of the unique data sets collected by EUI during the very first close perihelion RSWs, covering the period from 2022 March 2 to 2022 April 6 (Fig.\,\ref{fig:LTP6orbit.pdf}, reproduced from the ESA Solar Orbiter website\footnote{https://issues.cosmos.esa.int/solarorbiterwiki/display/SOSP}\label{SOCpublic}). As compared to Earth, Solar Orbiter observed from a perspective of increasing solar longitude (i.e.\ East to West), and transitioned from near-alignment with Earth on 2022 March 6, to perihelion on 2022 March 26, and then in a quadrature formation with the Earth, observing the Sun above the West limb, on 2022 March 29. During that period, the distance to the Sun from Solar Orbiter varied from 0.32\,au to 0.55\,au, closer to the Sun than any coronal EUV imager before. Given the variable angle with the Earth, the variable distance to the Sun, and the constraints of the low telemetry bandwidth, the instrument operations of the Remote Sensing Payload on Solar Orbiter was highly non-synoptic.  The aim of the paper is to guide the EUI data user through the unique but very variable EUI observations that were collected in the period 2022 March 2 to April 6. In Section \ref{section:SOOPs} we present the various EUI observation campaigns that were taken during the perihelion. In Section \ref{section:Highlights} we give an overview of the observational highlights that were completed in these campaigns.  After that, in Section \ref{section:Performance}, we describe for each EUI 
telescope the instrument performance. Finally, in Section \ref{section:Conclusions}, we give an outlook for upcoming orbits.

\section{From Science Goals to EUI Data sets}
\label{section:SOOPs}

As Solar Orbiter is a deep space, non-synoptic probe, the activity scheduling of its instruments is coordinated well in advance. This is done through Solar Orbiter Observing Plans (SOOPs), which bring a group of Solar Orbiter instruments in a specific mode to target a certain science goal at appropriate times during the orbit. These science goals have been generically described in \cite{2020A&A...642A...3Z} and updates are maintained on the ESA Solar Orbiter website.
% \footref{SOCpublic}.   fragile...

In order to document the intended purpose and context of the collected observations, we here summarise the SOOPs that the EUI instrument was involved in during the 2020 March 2 to April 6 period.  Grouping the SOOPs by science target and by increasing operational complexity, we arrive at three themes: (1) observing the Middle Corona off limb, (2) observing the building blocks of the solar atmosphere and, (3) making the connection from high resolution on-disk observations to in-situ measurements of the solar wind. The full technical details of the resulting EUI data sets can be found in Table \ref{datasetTable} in Appendix\,\ref{s-appendixA}, where entries are linked to corresponding SOOP names.

\subsection{Observing the off limb Middle Corona\label{sec:middle.corona}}

The first set of SOOPs focused on the off limb corona. Most of these SOOPs are led by the Solar Orbiter coronagraph Metis, which requires Sun center pointing. This makes the SOOPs operationally simpler as no last-minute pointing corrections are needed. For these SOOPs, EUI FSI images are prioritized due to the large overlap with Metis \citep{2020A&A...642A...6A} with Metis. Optional HRI images are necessarily pointed near disc center.

\begin{figure}[ht]
\centering
\includegraphics[width=0.48\textwidth]{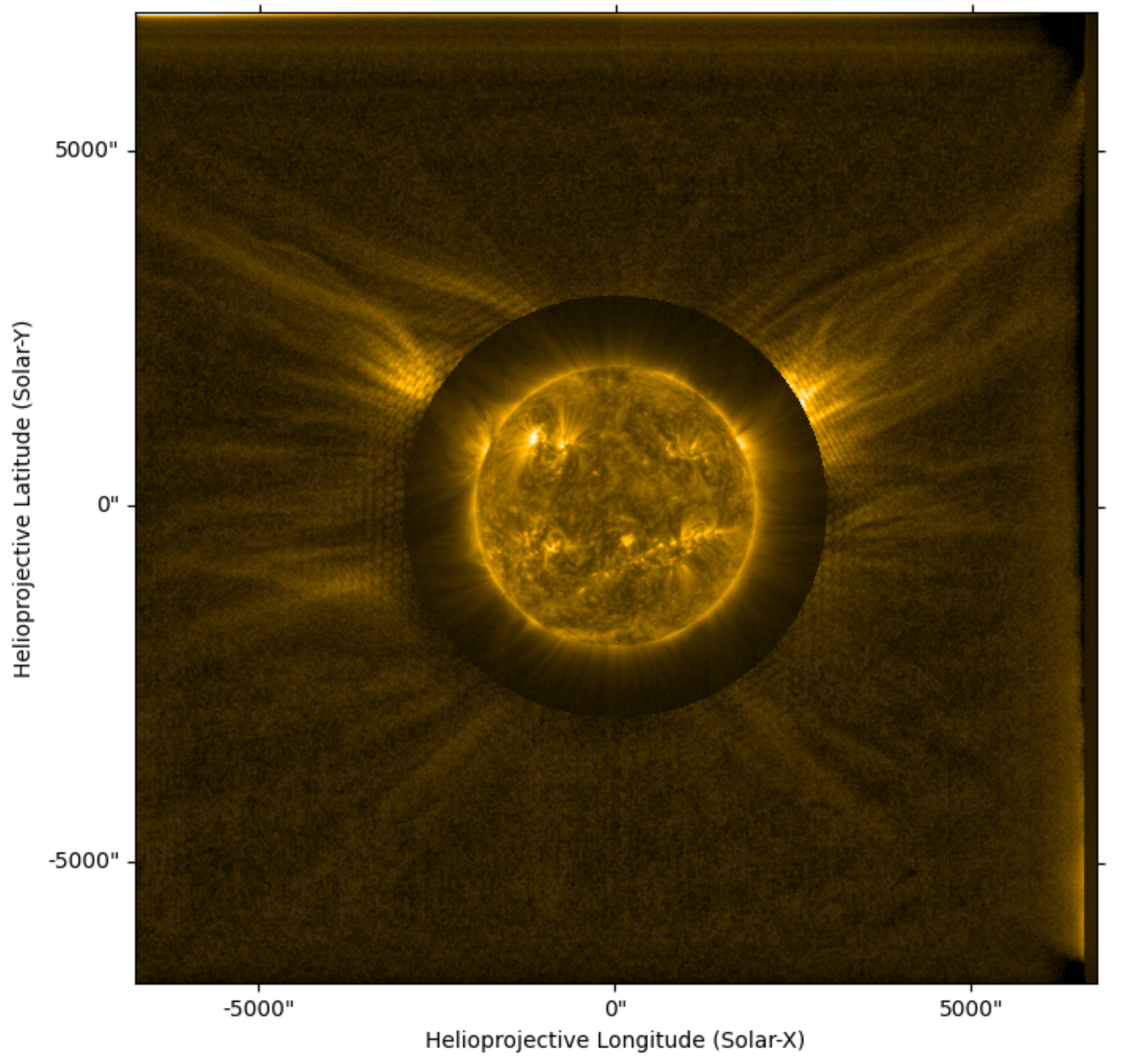}
\caption{FSI image taken in coronagraph mode (outer FOV) on 2022 March 7 at 16:00:05 UT, composited with a regular disk image (center) taken at 11:29:45 UT. The images were enhanced using the WOW algorithm ~\citep{Auchere2022} to reveal faint features. See Sect. \ref{sec:middle.corona}.}
\label{fig:fsi_helium_composite}
\end{figure}

% written by Frederic
The aim of the \hyperref[Coronal-He-Abundance]{L\_FULL\_HRES\_MCAD\_Coronal-He-Abundance} SOOP on 2022 March 7 was to support observations during the second launch of the Herschel sounding rocket, whose first flight provided the first helium abundance maps of the solar corona~\citep{moses2020}. The abundance is deduced from the ratio of the resonantly scattered intensities of neutral hydrogen and singly ionized helium, as imaged by two coronagraphs: SCORE~\citep{romoli2003} and HeCOR~\citep{auchere2007}. The method is model-dependent, and requires an independent knowledge of the temperature of the scattering ions. This can be constrained by simultaneous EUV observations, which was the purpose of the FSI observations at 17.4\,nm. In order to minimize stray-light at large distances from the solar limb, the instrument was used in coronagraph mode, with a movable disk masking direct sunlight~\citep[see ][]{auchere2005, Rochus2020}. One of the returned images is shown in Fig.\,\ref{fig:fsi_helium_composite}, composited with the closest in time disk image taken before the campaign. The sounding rocket payload failed, but the FSI data are still very useful to study the 17.4\,nm emission of the extended corona. The composite EUV image allows to link the magnetic structures on the disk (coronal holes, plumes, active regions) to magnetic field expansions in the extended corona, which appear always open but far from being simply radially aligned.

% orginally written by  Luciano
The \hyperref[Coronal-Dynamics]{L\_FULL\_HRES\_HCAD\_Coronal\-Dynamics} SOOP is designed to observe structures in the outer corona with the aim of linking them to the solar wind observed in-situ.  The SOOP was run twice, on 2022 March 22 (at 0.33\,au) and March 27 (at 0.32\,au) just after perihelion.  Given the outer corona focus, the main contribution of EUI was through FSI observations in both wavelengths which provide a large overlap with Metis. Nevertheless, HRI images of the quiet Sun at disc center were also taken at relative low cadence (30\,s to 60\,s). While these HRI images are perhaps not directly useful for the SOOP goal, they are unique observations as, given the close solar approach, they were the sharpest quiet Sun EUV images ever taken during the first perihelion passage.

% written by Cis
The goal of the \hyperref[Eruption-Watch]{L\_FULL\_HRES\_HCAD\_Eruption-Watch} SOOP is to observe eruptive events and to contribute to the understanding of Coronal Mass Ejections. The SOOP was carried out in two campaigns; the first one on 2022 March 22-23, and the second one between 2022 March 29-30  (close to Solar Orbiter in quadrature at the west side of Earth). There was long term monitoring with the FSI (every 6\,min) while the HRIs operated in 30\,min bursts.

\begin{figure}[ht]
\centering
\includegraphics[width=0.49\textwidth]{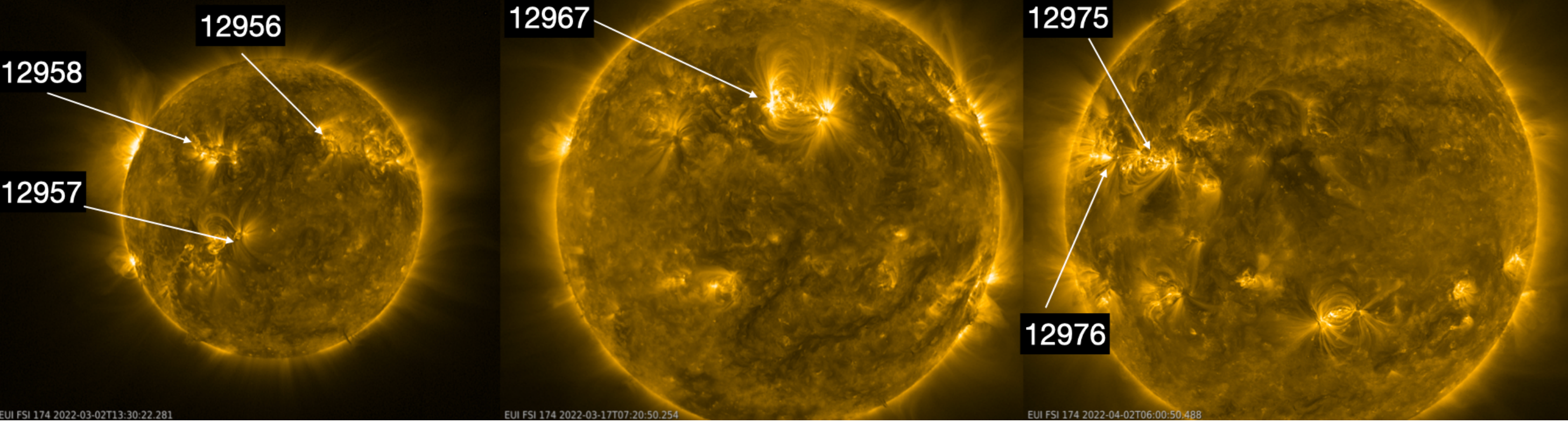}
\caption{Observed active regions on 2022 March 2 (left), March 17 (middle) and April 2 (right). See Sects. \ref{sec:building.blocks} and \ref{sec:in.situ.connection}.} \label{fig:ObservedActiveRegions.pdf}
\end{figure}

\subsection{Observing the building blocks of the on-disc solar atmosphere\label{sec:building.blocks}}

A second set of SOOPs targeted various features in the on-disc solar atmosphere. These SOOPs are typically led by SPICE \citep{SPICE} and/or EUI and require pointing the spacecraft to targeted features. Pointing corrections were made a few days in advance, commanded through the so-called "Pointing-Very Short Term Planning" (pVSTP, \cite{2020A&A...642A...3Z}) and based on Low Latency FSI images that are brought to the ground as a high priority. These are typically less than a day old.

The observation of small-scale EUV brightenings in the quiet Sun was an early success of EUI \citep{2021A&A...656L...4B}. Therefore, particular attention was paid to scheduling the \hyperref[Nanoflares]{R\_BOTH\_HRES\_HCAD\_Nanoflares} SOOP that is focused on surveying small impulsive events.
There are many open questions on their origin and properties. Are they also located in active regions and coronal holes? Similar events were observed in active regions by the Hi-C, but it is not clear yet if they are the same phenomenon. If this is the case, do they have the same properties everywhere?
Recent work suggests that a sub-population of the small-scale EUV brightenings does not reach the 1~MK
(Dolliou et al. 2022, (this volume, in revision); Huang et al. 2022, (this volume, in prep)).
Answering these questions will help us understand, for instance, their origin and relationship with the small scale magnetic field structuring and evolution.

The R\_BOTH\_HRES\_HCAD\_Nanoflares SOOP was scheduled several times near the Sun-Earth line (2022 March 7), at an angle with the Earth of roughly 30$^{\circ}$ and near quadrature (2022 March 30). The spacecraft pointed alternatively to an active region and to the quiet Sun.
This SOOP resulted in thousands of \hrieuv and \hrilya images obtained with a cadence of usually 3\,s (\hrieuv) and 5\,s (\hrilya) for a duration of approximately 30\,min (co-temporal in both channels). The \hrieuv initial plan was to run at 2s cadence, however the first run on 2022 March 6 failed and the cadence was decreased to 3\,s.
This very high cadence period was, in general, followed by longer periods at lower cadence. This choice came from the allocated telemetry and the need for long high cadence temporal sequences.
These campaigns were widely supported by other instruments on Solar Orbiter but also by IRIS \citep{DePontieu2014}, Hinode \citep{Kosugi2007}, PROBA2/LYRA \citep{2013SoPh..286...21D} and SDO/AIA. The latter instrument ran in a restricted configuration (subfield, few wavelengths) to image at an enhanced 6s  cadence in 4 wavelengths (13.1\,nm, 19.1\,nm, 17.1\,nm and 30.4\,nm channels).

The \hyperref[Polar-Observations]{R\_SMALL\_HRES\_MCAD\_Polar-Observations} SOOP, as described in \cite{2020A&A...642A...3Z}, aims at observing the polar magnetic fields by PHI \citep{PHI}. In this early phase of the mission however such observations are not ideal yet,  as the spacecraft  is still at low solar latitudes. Instead, the focus of the SOOP was to observe the polar coronal holes with the high resolution imagers of EUI. Such observations are timely as polar coronal holes will soon disappear with the rising solar cycle. This SOOP was carried out three times, once when Solar Orbiter was close to the Sun-Earth line (2022 March 6), once when Solar Orbiter was close to quadrature with Earth (2022 March 30) and once more when Solar Orbiter was at roughly 120$^{\circ}$ with the Earth (2022 April 4).
In the first two campaigns, the solar south pole coronal hole was the target  but in the last campaign the solar north pole coronal hole was more visible. In particular the 2022 March 30 observation of the south pole returned the `best ever' EUV images of a polar coronal hole as they were taken from 0.33\,au with an imaging cadence of 3\,s.

% written by Cis
The \hyperref[BrightPoints]{R\_BOTH\_HRES\_MCAD\_Bright-Points} SOOP is SPICE-focused and aims at observing coronal bright points.  This SOOP was carried out between 2022 March 8 08:10 and 14:10. During that time, FSI operated with a relatively high time cadence (5\,min).  \hrieuv and \hrilya observed during 2 hours at 1 minute cadence. The EUI telescopes pointed at disc centre (quiet Sun) and bright points were observed.

% written by Louise, Miho
The \hyperref[AR-Long-Term]{R\_SMALL\_MRES\_MCAD\_AR-Long-Term} was carried out between 2022 March 31 and 2022 April 4. The goal was to track the decay phase of an active region. Two good candidates appeared on the solar disk a few days before the SOOP started: NOAA AR\footnote{Active region numbering by https://www.swpc.noaa.gov} 12975 and NOAA AR 12976 (see Fig.\,\ref{fig:ObservedActiveRegions.pdf}). Comparing both regions, the leading polarity of NOAA AR 12976 provided a good target and the EUI FOV was chosen centered on it.  There was long term monitoring with FSI (every 30\,min) and burst image sequences with both HRIs, typically with a time cadence of 10\,s and lasting 47\,min. The regions observed during these few days produced several flares, including an M-class flare on 2022 April 2 that is discussed further below.

% written by Emil
On 2022 March 7, Solar Orbiter crossed the Sun-Earth line, at a distance of 0.49\,au, allowing for  cross-calibration with similar Earth-bound instruments.
For a complete inter-comparison, a full range of scenes (quiet Sun area, an active region or a coronal hole) was to be targeted within the small FOV of PHI/HRT, \hrieuv, \hrilya and SPICE. However, by having Solar Orbiter point in a 5x5 pattern, these  high resolution telescopes could however make a \hyperref[Full Disc Mosaic]{Full Disc Mosaic} of the whole solar disc, thereby avoiding  upfront guessing the position of the various scenes.
Solar Orbiter followed the 5x5 pointing pattern from north-east (top left) to south-west (bottom right) in columns.
% The time needed to re-point Solar Orbiter from one orientation to the next (slew time) is 5 min.
% The time needed to keep 1 pointing (dwell time) is also 5 min and is driven by the time needed for SPICE rasters. This resulted in images between the dwells being taken 10 min apart in vertical direction and 50 minutes apart in horizontal direction.
Images in subsequent pointing positions are 10\,min apart in the vertical direction and 50\,min apart in the horizontal direction.
To ensure a maximum overlap between image panels, the \hrieuv images were commanded to be 2368x2368 in size.
%which meant that the field of view was not fully illuminated, with the intensity dropping off towards the corners and edges over a distance of ~30 pixels.
On average, the images between dwells overlapped by ~600 pixels.
The \hrieuv telescope took high-gain and low-gain image pairs every 30\,s within each dwell period, resulting in 9 such image pairs per pointing position. The high- and low-gain images were taken 5\,s apart and calibrated and combined on-ground into high dynamic range images (15-bit).  To create a high resolution mosaic of the full Sun, these high dynamic range images from each dwell position were aligned and stitched together using affine image transformation making use of the spacecraft attitude information available in the source FITS files. The resulting panels were then blended together manually in photo editing software, minimizing image artifacts in the mosaic caused by changing views in neighboring panels due to dynamic events and solar rotation. The panels were blended together preferentially in quiet Sun areas, avoiding the faster changing active regions where possible.
% The contrast in the resulting mosaic was increased with unsharp masking, and to increase the field of view of the off-disc solar corona, the entire mosaic was blended on top of an EUI/FSI 17.4~nm image that was exposed during the same dwell times.
The final mosaic has more than 83 million pixels, making it the highest resolution image of the Sun’s full disc and corona ever taken. An interactive version of the image can be found in \cite{Kraaikamp2022PosterBelfast}.

\begin{figure*}[ht]
\centering
\includegraphics[width=0.99\textwidth]{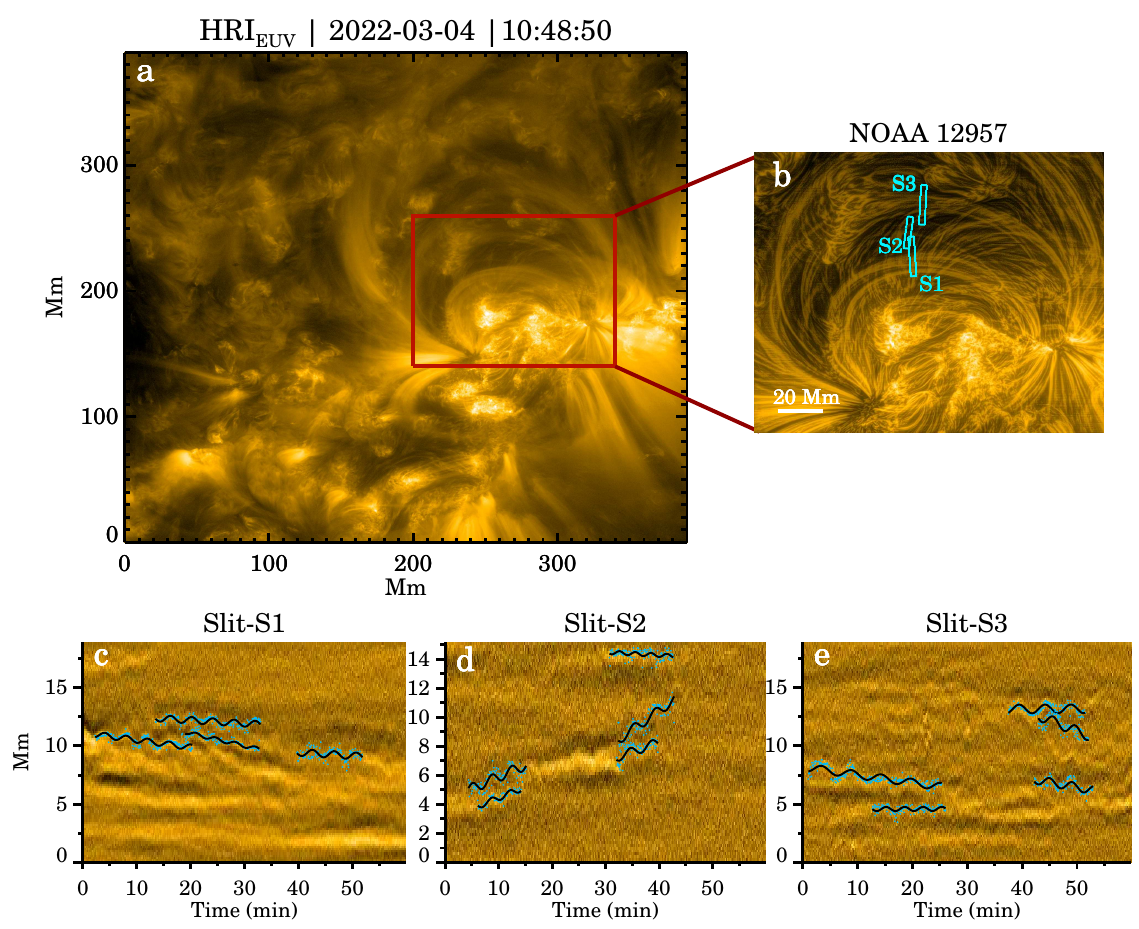}
\caption{Example of decayless kink waves observed in AR12957 on 2022 March 04. Panel b shows a zoom into the loops in panel a. The boxes S1 to S3 mark slits along which the temporal evolution is shown in the form of space-time diagrams in panels c to e. To enhance the appearance of the oscillating threads in these space-time diagrams, a smooth version of the map (boxcar-smoothed in the vertical direction) is subtracted from each original map. The black lines represent fits for the oscillation of the loops. See Sect.~\ref{sec:decayless} for details.} \label{fig:AR19257_March3.pn}
\end{figure*}

\subsection{Making the connection from high resolution on disk to in-situ\label{sec:in.situ.connection}}  %%%%%%%%%%%%%%%%%%%%%%%%%%%%%%%%%

A third set of SOOPs aimed at finding the connection between the smallest features imaged on disc to the corresponding in-situ measurements of the out flowing solar wind.

% originally written by Luciano Rodriguez
The goal of the \hyperref[Connection-Mosaic]{L\_SMALL\_MRES\_MCAD\_Connection\-Mosaic} SOOP is to identify, with SPICE and the high resolution imagers of EUI and PHI, the connection point on the solar disc for the solar wind observed in situ.  In order to increase the probability of successfully catching the connection point, this SOOP is implemented in combination with a mosaic of spacecraft pointings. The SOOP was run twice, once when Solar Orbiter was near the Sun-Earth line (2022 March 2-3, at 0.56\,au) and once when Solar Orbiter was in quadrature with Earth (2022 March 30, at 0.33\,au). In the first instance, the spacecraft made a mosaic of 3 vertically aligned pointings, in the second instance, a mosaic of 3x2 pointings was made.
Each of these pointings was maintained for several hours. It was later discovered that solar rotation was not correctly compensated, so the EUI FOV slightly shifts over this duration. The location of the mosaics was decided a few days in advance with the help of various models and tools \citep{2020A&A...642A...2R}.
During the first instance of this SOOP, EUI observed an M2 flare in the high resolution FOV, see section \ref{sec:M2.flare}.

% written by David Long
The \hyperref[Slow-Wind-Connection]{L\_SMALL\_HRES\_HCAD\_Slow-Wind-Connection} SOOP (Yardley, S. et al, 2022, this issue, in preparation) was designed to combine the remote-sensing and in-situ capabilities of Solar Orbiter, observing the source of solar wind connected to the spacecraft and then detecting the plasma released from the Sun as it passed over the spacecraft several days later. Although the primary instruments used in the SOOP are SPICE and SWA \citep{SWA}, the EUI HRI observations provided high resolution context images. Due to the need to identify where material passing over the spacecraft was originally ejected from the Sun, the SOOP coordinator relied heavily on the connectivity tool developed by IRAP \citep{2020A&A...642A...2R} to identify the origin of the magnetic field predicted to be connected to Solar Orbiter during the observing window. For the first observing window (2022 March 3-6), this was the boundary between NOAA AR~12957 and a nearby equatorial coronal hole (Fig.\,\ref{fig:ObservedActiveRegions.pdf}). For the second window (2022 March 17-22), two different targets were selected due to a change in the connectivity of the spacecraft with respect to the solar magnetic field. The boundary of the southern polar coronal hole was selected as the target for the first part of the observing window, with the positive polarity of NOAA AR~12967 in the northern hemisphere selected as the target for the second part of the window.

% written by Cis
The \hyperref[CH-Boundary-Expansion]{L\_BOTH\_HRES\_LCAD\_CH-Boundary-Expansion} SOOP is similar to L\_SMALL\_HRES\_HCAD\_Slow-Wind-Connection discussed above, but it specifically aims to study coronal holes boundaries as possible sources of the slow solar wind. The SOOP was active between 2022 March 25 19:40  and 2022 March 27 00:00. FSI acquired 17.4\,nm and 30.4\,nm images at 10\,min cadence. The quiet Sun at disc centre was observed in both campaigns.

\section{Observational highlights}
\label{section:Highlights}

In the previous section, the EUI observations of the 2022 March 2--April 6 period were presented from a science planning perspective.
Solar activity however seldom follows the science plan so we additionally review the actual observational highlights here that were identified so far
 in the collected data. While this 35-day period is longer than a solar rotation period, its sub-solar point ranged from 64$^{\circ}$ Carrington longitude in the beginning of the period to 95$^{\circ}$ Carrington longitude at the end of the period, due to the intrinsic motion of the spacecraft over the same period. The longitudinal angle  Earth-Sun-spacecraft ranged from -8$^{\circ}$ to 122$^{\circ}$. In what follows we use EUI Data Release 5 \citep{euidatarelease5}.

\subsection{Active region dynamics}      %%%%%%%%%%%%%%%%%%%%%%%%%%%%%%%%%%%%%%%%%%%%%%%%%%%%
\label{ActiveRegionDynamics}

Several active regions (see Fig.\,\ref{fig:ObservedActiveRegions.pdf}) were observed in the \hrieuv and \hrilya FOVs with imaging cadences sometimes as fast as 3\,s. Below we highlight a sample of some particular dynamics..

%%%%%%%%%% March 3, NOAA 2957} %%%%%%%%%%
\subsubsection{Decayless kink oscillations}\label{sec:decayless}
NOAA AR 12957 was observed first as part of the mosaic pattern of \hyperref[Connection-Mosaic]{L\_SMALL\_MRES\_MCAD\_Connection\-Mosaic} (2022 March 2, 3) and then as part of the daily high resolution bursts of \hyperref[Slow-Wind-Connection]{L\_SMALL\_HRES\_HCAD\_Slow-Wind-Connection} on 2022 March 3, 4 and 5. At this time, Solar Orbiter was at 0.544\,au from the Sun, resulting in an  \hrieuv pixel footprint of $194~km$ on the Sun.
The core of the active region showed a myriad of counter-streaming loops, of which some exhibited decayless kink oscillations (Fig.\,\ref{fig:AR19257_March3.pn}). This event is of particular interest due to the fact that these oscillating loops are rooted inside sunspots which are generally devoid of supergranular flows, the commonly assumed driver of such decayless kink oscillations. Moreover, these decayless oscillations were only observed during specific time intervals although the loop environment remained more-or-less similar throughout. Further details on the magnetic configuration of those loop footpoints, as well as the existence of other possible wave drivers are presented in \citet{Mandal2022}.

%%%%%%%%%% March 17 Saint Patrick AR, braiding loops  %%%%%%%%%%
\subsubsection{Braiding loops\label{sec:braiding.loops}}
\begin{figure}[ht]
\centering
\includegraphics[width=0.49\textwidth]{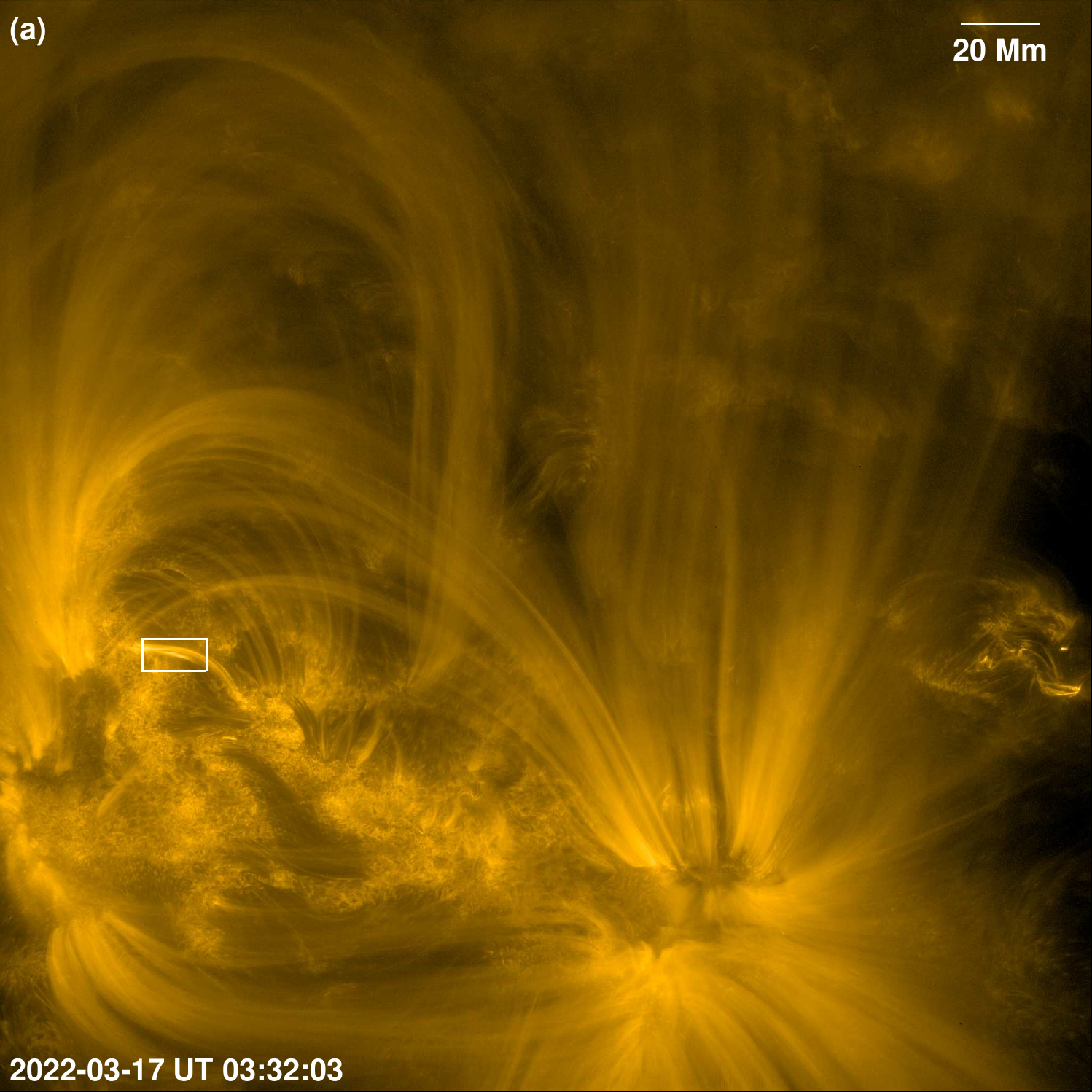}
\includegraphics[width=0.49\textwidth]{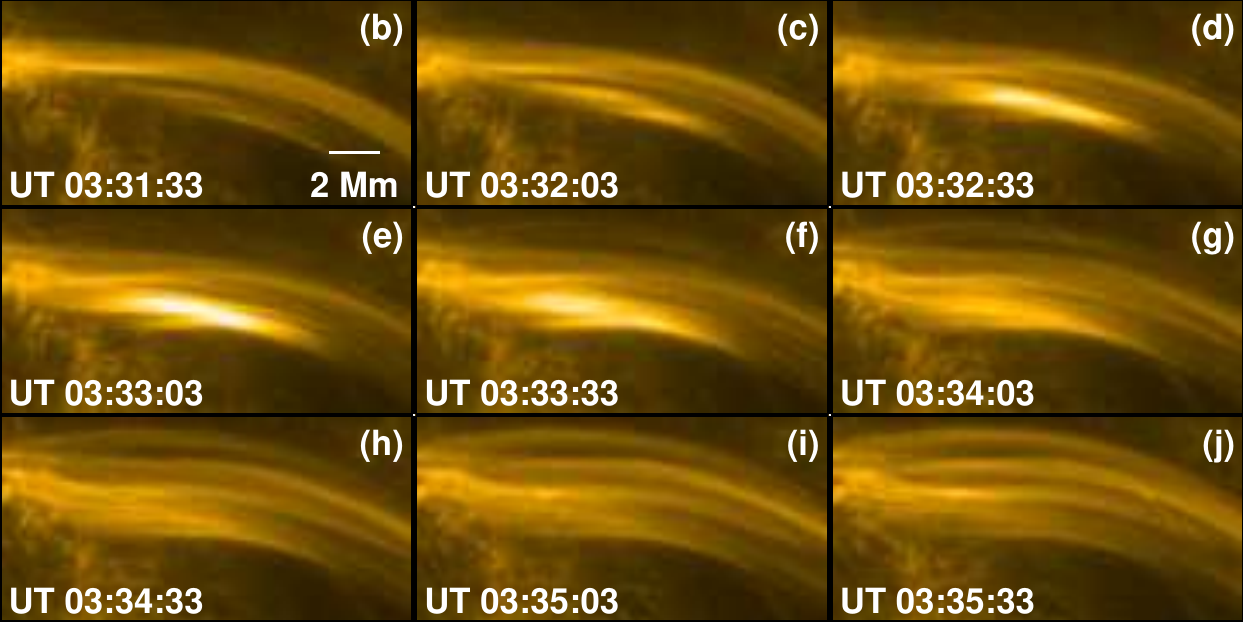}
\caption{Example of a relaxation of braided coronal loops observed on 2022 March 17. Panels (b) to (j) display a zoom into the loop in panel (a) (white box) and show the evolution of the untangling within the loop. See Sect.~\ref{sec:braiding.loops} for details.} \label{fig:braid1}
\end{figure}

As a part of the \hyperref[Nanoflares]{R\_BOTH\_HRES\_HCAD\_Nanoflares} SOOP on 2022 March 17, \hrieuv observed active region AR12965 at a cadence of 3\,s. These are among the first highest cadence EUV images of an active region ever observed. During this period, Solar Orbiter was at a distance of 0.38\,au. Thus the 2-pixel footprint of \hrieuv was about 270\,km on the Sun. An overview of the observed active region is displayed in Fig.\,\ref{fig:braid1}a. These high-resolution, high-cadence observations of this active region revealed a number of impulsive EUV brightenings on timescales of a few minutes or less. A closer look at some of the brightenings revealed that they are associated with untangling of braided coronal strands or loops. Most of these events are observed in shorter, low-lying loop features. \hrieuv also observed untangling of coronal braids in a more conventional loop system. A sequence of this untangling of braided loops is shown in Fig.\,\ref{fig:braid1}b--j. More details on examples of braided structures observed by \hrieuv and the implications for coronal heating are discussed in \citet{chitta2022}.

%%%%%%%%%% April 1, AR12975 and AR12976 : Highly dynamic cooler loops}
\subsubsection{Highly dynamic cooler loops\label{sec:dynamic.loops}}

\begin{figure}[ht]
\centering
\includegraphics[width=0.49\textwidth]{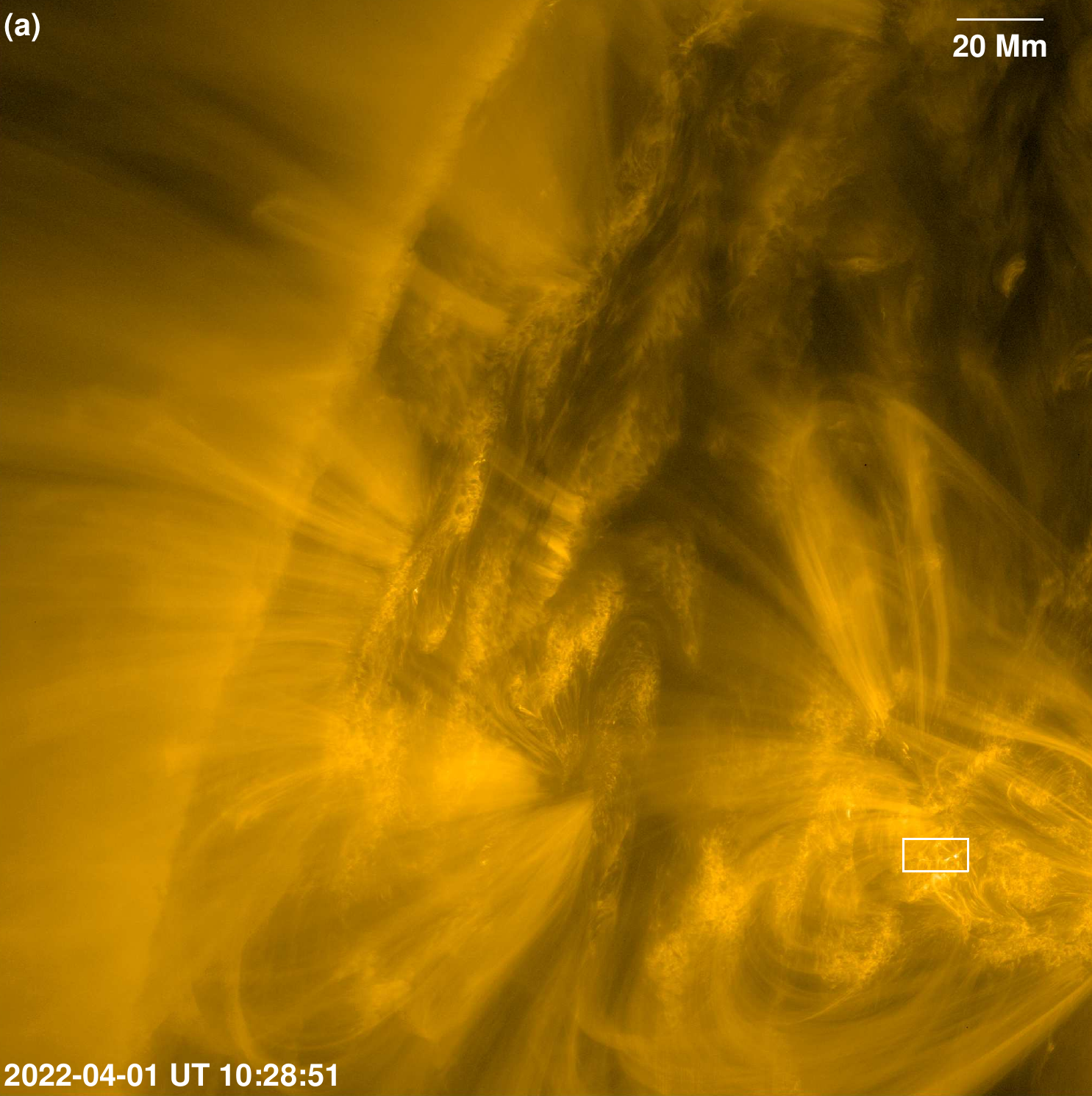}
\includegraphics[width=0.49\textwidth]{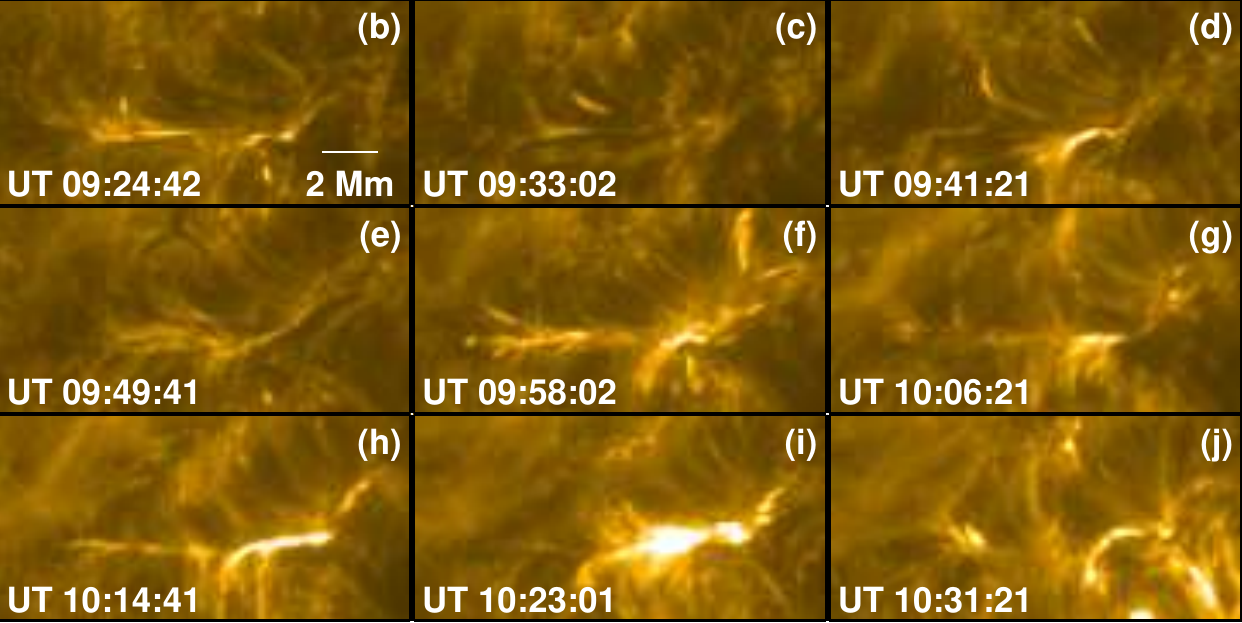}
\caption{Example of a highly dynamic cool loop system in the core of active region AR12975 observed by \hrieuv on 2022 April 1. See Sect. \ref{sec:dynamic.loops}} \label{fig:fibril1}
\end{figure}

The region of interest of R\_SMALL\_MRES\_MCAD\_AR-Long-Term on 2022 April 1 was on the east limb of the Sun from the vantage point of Solar Orbiter, which means that the same region appeared in the western hemisphere from the viewpoint of Earth. Fig.\,\ref{fig:fibril1}a shows the full FOV of \hrieuv, with NOAA AR 12975 on the western side and NOAA AR 12976 on the eastern side. Besides imaging instances of coronal braids in low-lying loop systems in both these active regions \citep[see discussion in][]{chitta2022}, \hrieuv captured a highly dynamical system of cooler loops that appear darker in EUV due to absorption. These dynamic loops were observed in the core of AR12975 (see white box in Fig.\,\ref{fig:fibril1}). These are likely related to chromospheric arch-filament systems associated with emerging flux regions \citep[][]{2015LRSP...12....1V}. Individual strands or loops within this system exhibited intermittent brightenings in EUV. In addition, there are also repeated compact brightenings on one end of this feature. The morphology of these compact EUV brightenings appear akin to the transition region ultraviolet bursts that are often observed in emerging flux regions \citep[][]{2018SSRv..214..120Y}. The evolution of this region over a period of 1\,hour is displayed in a sequence of images in Fig.\,\ref{fig:fibril1}b--j.

% \url{https://chat.observatory.be/eui/pl/pyxpr41cuidsfb81j3wjhxo15w}
%%%%%%%%%%

%%%%%%%%%% April 1, AR12975 and AR12976 : Highly dynamic cooler loops}
\subsubsection{Brightening on border of dark material\label{sec:bright.next.to.dark}}

\begin{figure}[ht]
\centering
\includegraphics[width=0.45\textwidth]{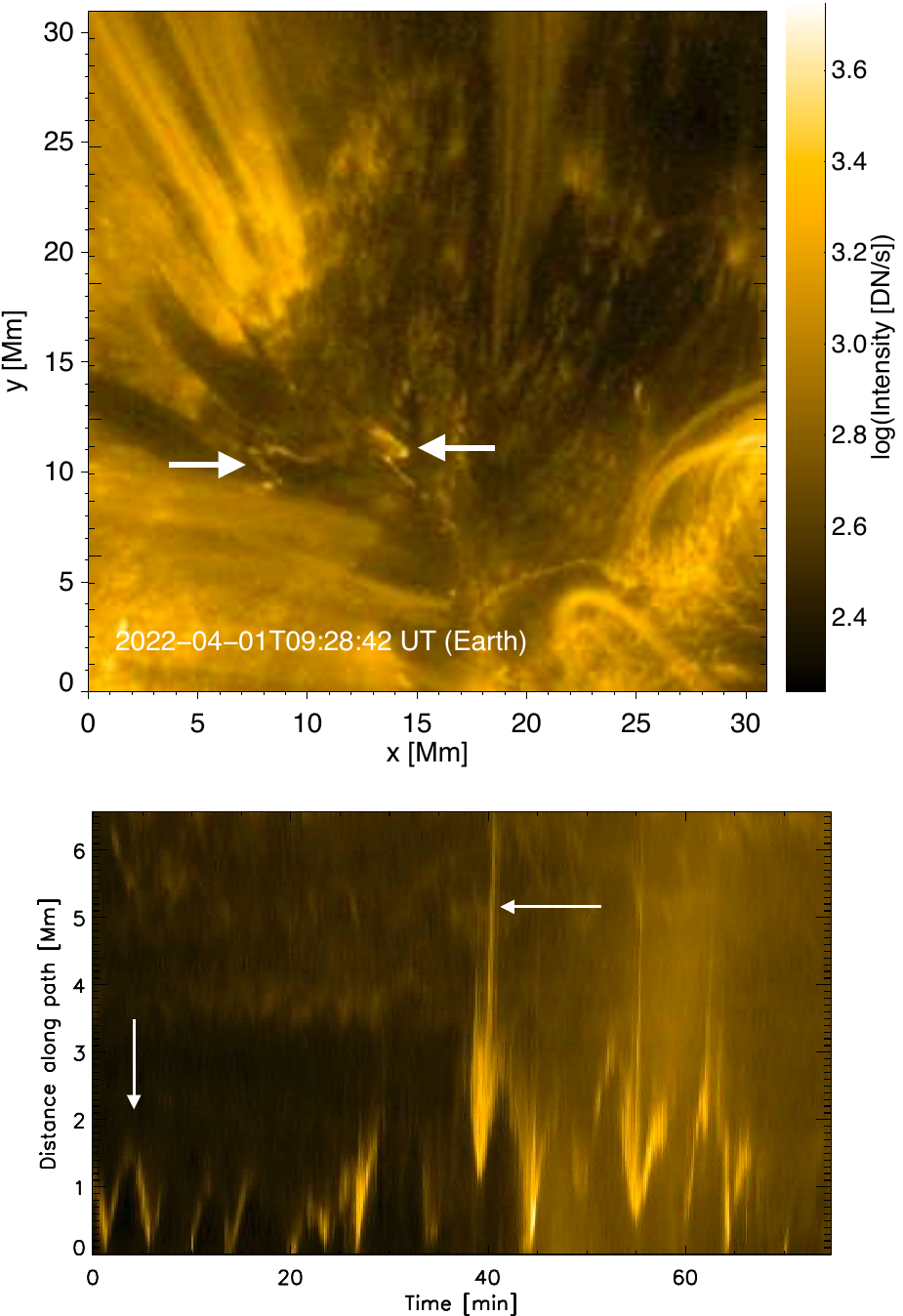}
\caption{\hrieuv observations of the brightening on a border of dark material in an active region on 2022 April 1. The white arrows in  the top panel mark the brightening location. The  bottom panel shows a time-distance map along one of the jets observed propagating from the bright structure. Note the existence of 2 kinds of brightening events, the bright edge of the dark structure oscillating up and down, and jets propagating upwards (both indicated by the arrows). See Sect. \ref{sec:bright.next.to.dark}.
% (See also movie "movie\_brightenings.mp4")
} \label{fig:brightenings}
\end{figure}

During the R\_SMALL\_MRES\_MCAD\_AR-Long-Term SOOP on
2022 April 1, \hrieuv observed brightening events at the top of dark jet-like structures
%Brightening events at the top of dark jet-like structures were observed on 2022 April 01 at AR NOAA 12975  by \hrieuv during the entire sequence 09:19:15 (+@6m25s)  UTC to 10:32:35 (+@6m25s) UTC (Fig.\,\ref{fig:brightenings}) during the campaign R\_SMALL\_MRES\_MCAD\_AR-Long-Term.
The angular separation between the Earth and Solar Orbiter was 104$^\circ$ which allowed for simultaneous observation from SDO/AIA.  Solar Orbiter was 0.35\,au from the Sun, resulting in a spatial resolution of \hrieuv images (two pixel size) of 248\,km.
The dark jet-like structures appear to be the so-called light walls \citep{Hou_2016AA...589L...7H} or fan-shaped jets \citep[][]{2016A&A...590A..57R},
%or peacock tails
 which are field-aligned long chromospheric jets thought to be produced by magnetic reconnection in the photosphere \citep{Bharti_2015MNRAS.452L..16B,Bai_2019}. Two kinds of brightening events can be distinguished (see time-distance map in Fig.\,\ref{fig:brightenings}). The first kind is continuously present at the chromosphere-corona interface of the jets and is seen to oscillate up and down with a ballistic motion, strongly suggesting that this corresponds to the upward motion of the transition region observed by \hrieuv (17.4\,nm). Its narrow thickness (as small as 200\,km or less) and strong brightness is probably due in part to the passage from high to low density and cool to hot plasma, to which \hrieuv is particularly sensitive. The second kind of brightening is far more impulsive, with life times on the order of 10~s or less, and appears on top of the first kind as perturbances propagating upwards at speeds of $\approx100~$km~s$^{-1}$. Both brightening events are therefore likely due to slow mode shocks generated from the reconnection events lower down, first propagating in the chromosphere and then into the corona at the tube speed. The second kind is seen to originate when the dark structure is at the lower end of the oscillation range, as expected from chromospheric shocks leading to spicule-like events \citep{Heggland_2007ApJ...666.1277H,Heggland_2011ApJ...743..142H}.

%Patrick: I have tweaked the text above adding the 2 kinds of brightening events that we see.
% 21/09/2022: i have added a panel to the figure where we can clearly see the 2 kinds of perturbations. The speeds indicate that both are due to shocks, the first propagating in the chromosphere and pushing the TR up, and the second propagating in the corona.
%Victim: Patrick/Pradeep/Krzystof
%https://chat.observatory.be/eui/pl/uznawugdyt8o8q7goqswxipf8a \newline
%https://chat.observatory.be/eui/pl/ednfzz4fpjd38njzwhmkqt1kko \newline
%https://chat.observatory.be/eui/pl/ednfzz4fpjd38njzwhmkqt1kko
%%%%%%%%%%

\subsubsection{Coronal rain}\label{sec:rain}
\begin{figure}
\centering
\includegraphics[width=0.47\textwidth]{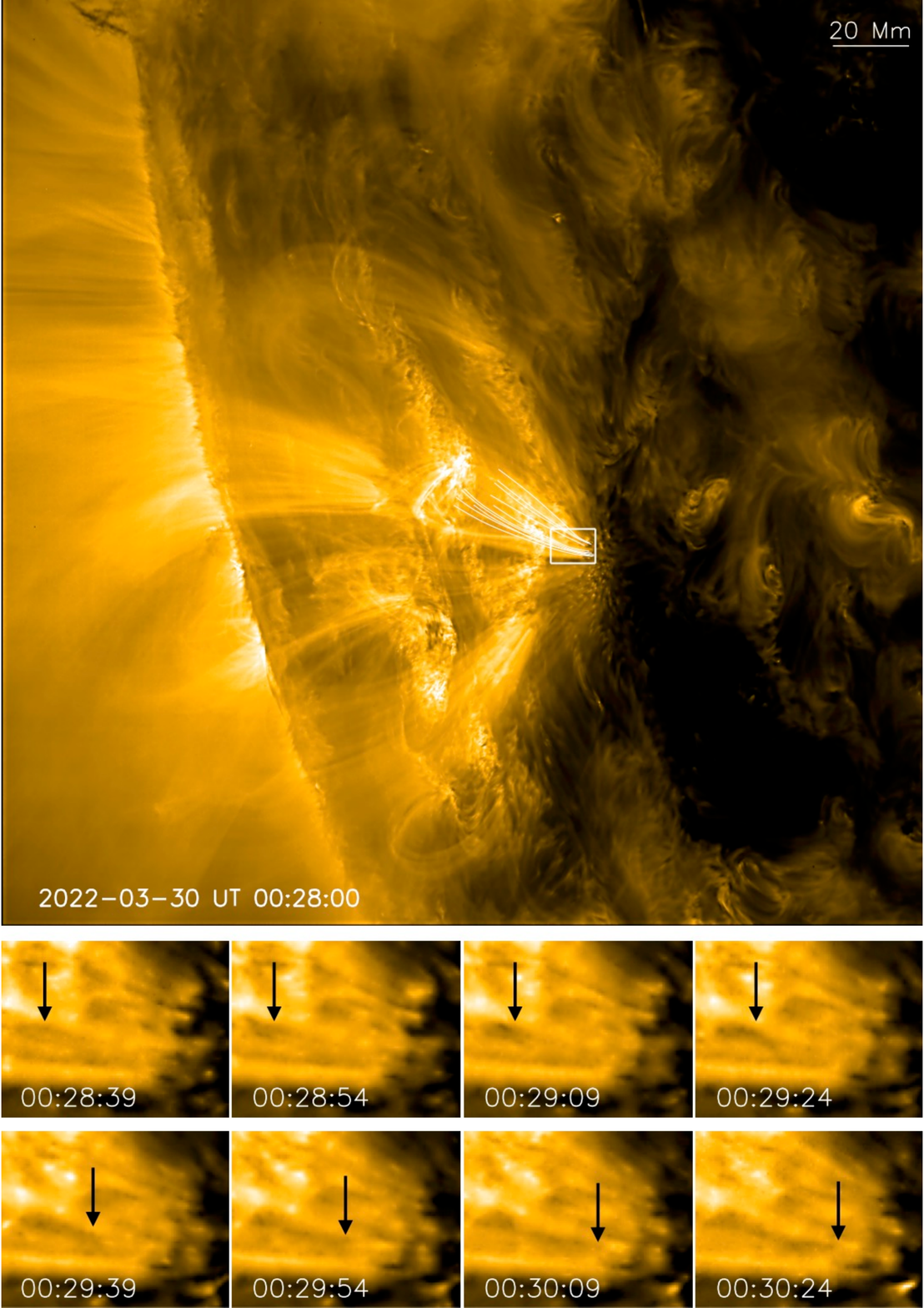}
\caption{Top: Field-of-view of the 2022 March 30 dataset observed by \hrieuv showing an active region at the South-East limb. The solid white curves follow several trajectories of coronal rain clumps seen in EUV absorption. Bottom panels: Snapshots at various instances of a coronal rain shower, corresponding to the small white rectangle shown in the top panel. The black arrows indicate the head of the coronal rain shower. See Sect. \ref{sec:rain}.} \label{fig:solorain}
\end{figure}

%Patrick
Coronal rain appears ubiquitous on-disk in AR NOAA 2975 and 2976 observed on 2022 March 30, April 1 and 2 (see Fig.\ref{fig:solorain}). In \hrieuv, coronal rain can be clearly distinguished in EUV absorption by its dynamics (with velocities close to 100\,km~s$^{-1}$ in the plane-of-the-sky) and its clumpy and multi-stranded morphology \citep{Antolin_2015ApJ...806...81A, Antolin_2020PPCF...62a4016A}. At the spatial resolution of $\approx$250\,km, individual coronal rain clumps only a few pixels wide can be observed. Their morphology is strongly reminiscent of H$\alpha$ high-resolution observations \citep{Antolin_Rouppe_2012ApJ...745..152A}, a similarity that has been predicted but so far only observed at larger loop scales \citep{Anzer_Heinzel_2005ApJ...622..714A,YangB_2021ApJ...921L..33Y}. Coronal rain showers (composed of clumps) can be observed in loop bundles rooted to moss, but both clumps and showers (despite the large shower widths above 10~Mm) appear mostly unresolved in AIA passbands. This picture therefore constitutes a major difference to previous EUV observations of active regions. The on-disk observation at high resolution provides a connection to the chromosphere (and photosphere with PHI), thus providing a unique insight of the heating events at the footpoints that lead to thermal non-equilibrium and instability associated with this phenomenon \citep{Antolin_Froment_10.3389/fspas.2022.820116}. A full paper reporting coronal rain observed with \hrieuv is available in Antolin et al. 2023 (this volume, in prep.).

\subsubsection{Hints of torsional Alfv\'en waves in twisted coronal loops\label{sec:torsional.waves}}
\begin{figure}
\centering
\includegraphics[width=0.48\textwidth]{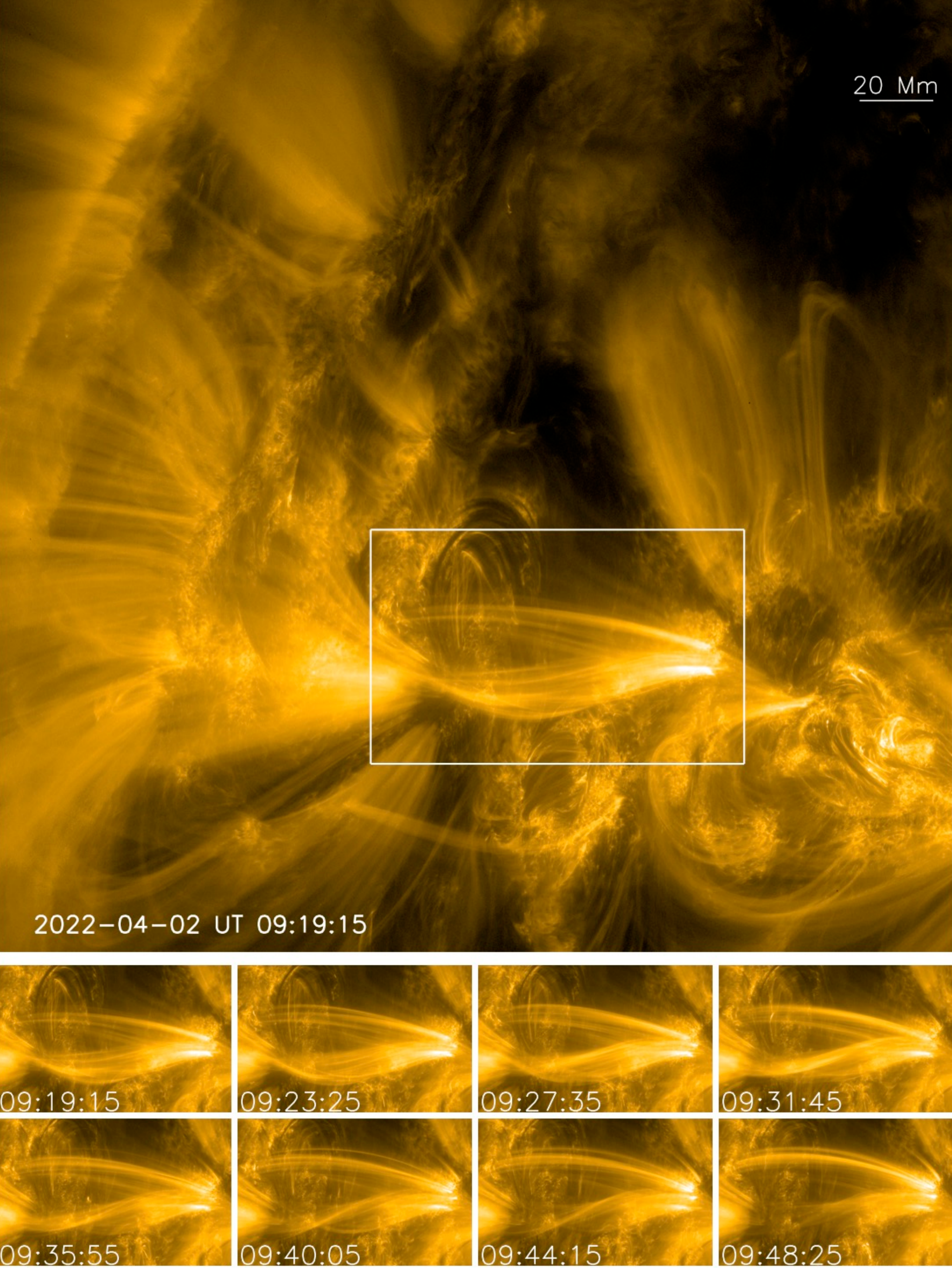}
\caption{Top: Field-of-view of the 2022 April 2 dataset observed by \hrieuv showing an active region at the South-East limb. The solid white rectangle shows a twisted coronal loop where hints of torsional Alfv\'en waves may be observed. Bottom panels: snapshots over 30\,min evolution of the loop observed in the field-of-view corresponding to the white rectangle. Note the change in orientation of the twisted EUV strands. See Sect. \ref{sec:torsional.waves}} \label{fig:solotaw}
\end{figure}

%Patrick
In AR NOAA 2975 observed on 2022 April 2nd twisted, intertwined coronal strands can be observed in the plane-of-the-sky, appearing and disappearing on a timescale of $5-10$\,min. The strands disappear by the end of the sequence with hints of untwisting, ending in coronal rain falling onto bright footpoints rooted in moss. The entire event is strongly reminiscent of the coronal loop model of \citet{DiazSuarez_2021ApJ...922L..26D} in which, torsional Alfv\'en waves propagate along a twisted flux tube. The radially varying magnetic field due to the twist provides an Alfv\'en continuum that allows phase mixing to happen. The Kelvin-Helmholtz instability is then generated due to the velocity shear at the phase mixing layers, which leads to compression of the plasma and the generation of coronal strands that follow the twisted flux tube  \citep[a process also observed for kink waves,][]{Antolin_2014ApJ...787L..22A}. The time-scale of appearance/disappearance of strands, their morphology and change in orientation during the oscillation is seen to match the one observed in this event with \hrieuv (see Fig.\,\ref{fig:solotaw}).

\subsubsection{Large-scale reconfiguration of coronal loop sub-structure}\label{sec:reconfig}
\begin{figure}[ht]
\centering
\includegraphics[width=0.48\textwidth]{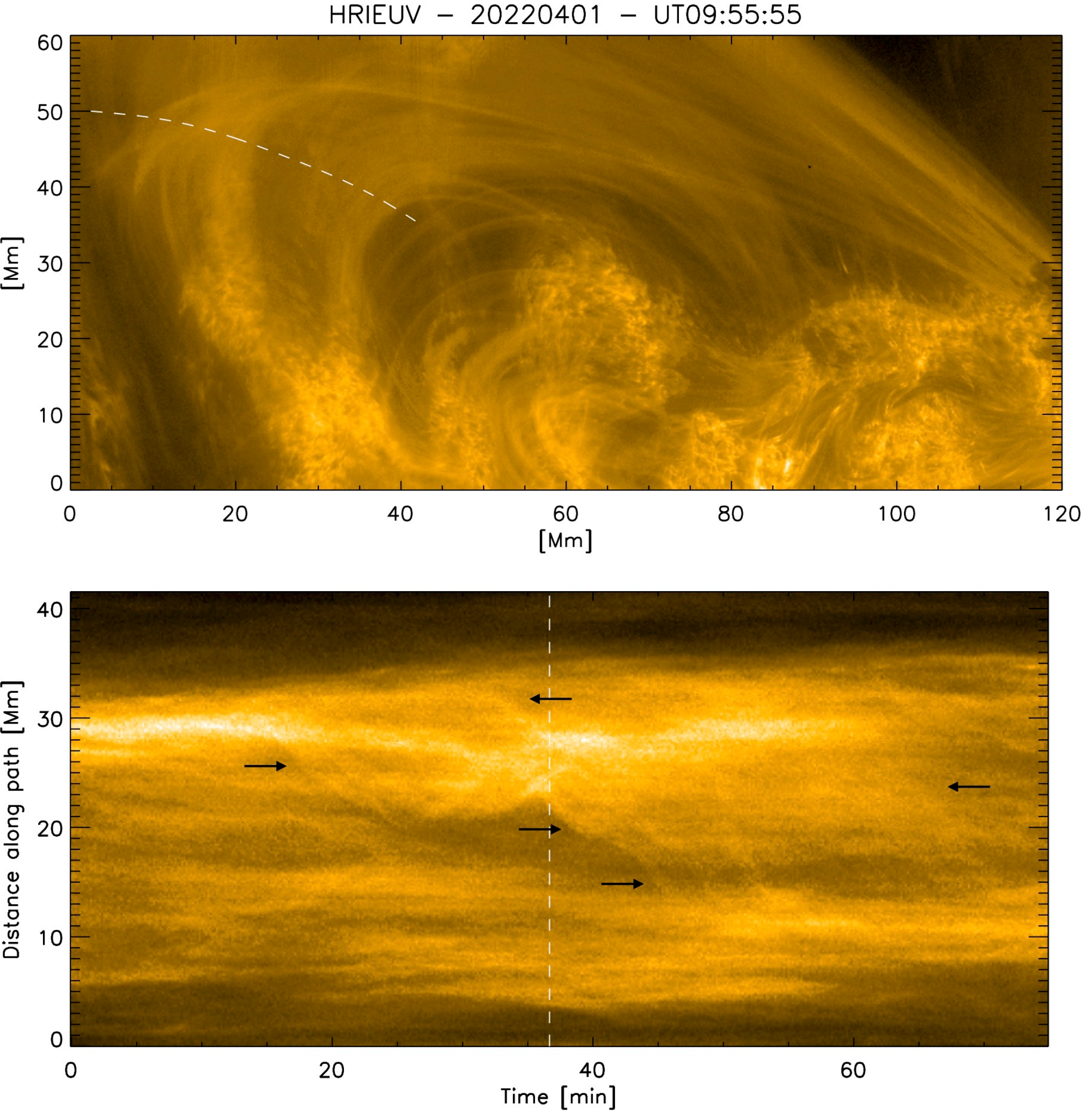}
\caption{Top: A sub-field of the 2022 April 1 dataset observed by \hrieuv (located on the bottom-right of Fig.\,\ref{fig:fibril1}). A loop bundle is seen (inverted in this figure in order to have the apex on top), where a large scale reconfiguration is observed. Bottom: Time-distance plot across the apex of the loop (dashed white curve on top panel), revealing a downward (inward) motion of many loops (akin to contraction), indicated by the black arrows, accompanied by transverse oscillations. The time of the snapshot on the top panel corresponds to the vertical white dashed line. See Sect. \ref{sec:reconfig}.} \label{fig:solorecon}
\end{figure}

%Patrick
In AR NOAA 2975 observed on 2022 April 1, a large scale loop bundle rooted in moss is composed of various strands. Without the presence of any flare in the vicinity, the strands undergo a coherent reconfiguration akin to contraction following a flare (see Fig.\,\ref{fig:solorecon}). The strands also exhibit continuous kink motions during the global contracting motion. The overall event is accompanied by coronal rain.

%\subsubsection{Longitudinal propagating disturbances}
% Patrick
%it may be worth to write a section on this. The intensity variation associated with p-mode or chromospheric wave leakage into the corona is usually reported to be a few percent. However, SolO data suggests far stronger intensity variation, which would result in far stronger wave amplitudes and therefore stronger wave energy fluxes. Need to check!

\subsubsection{Coronal strands associated with coronal rain\label{sec:strands.and.rain}}
\begin{figure}[ht]
\centering
\includegraphics[width=0.48\textwidth]{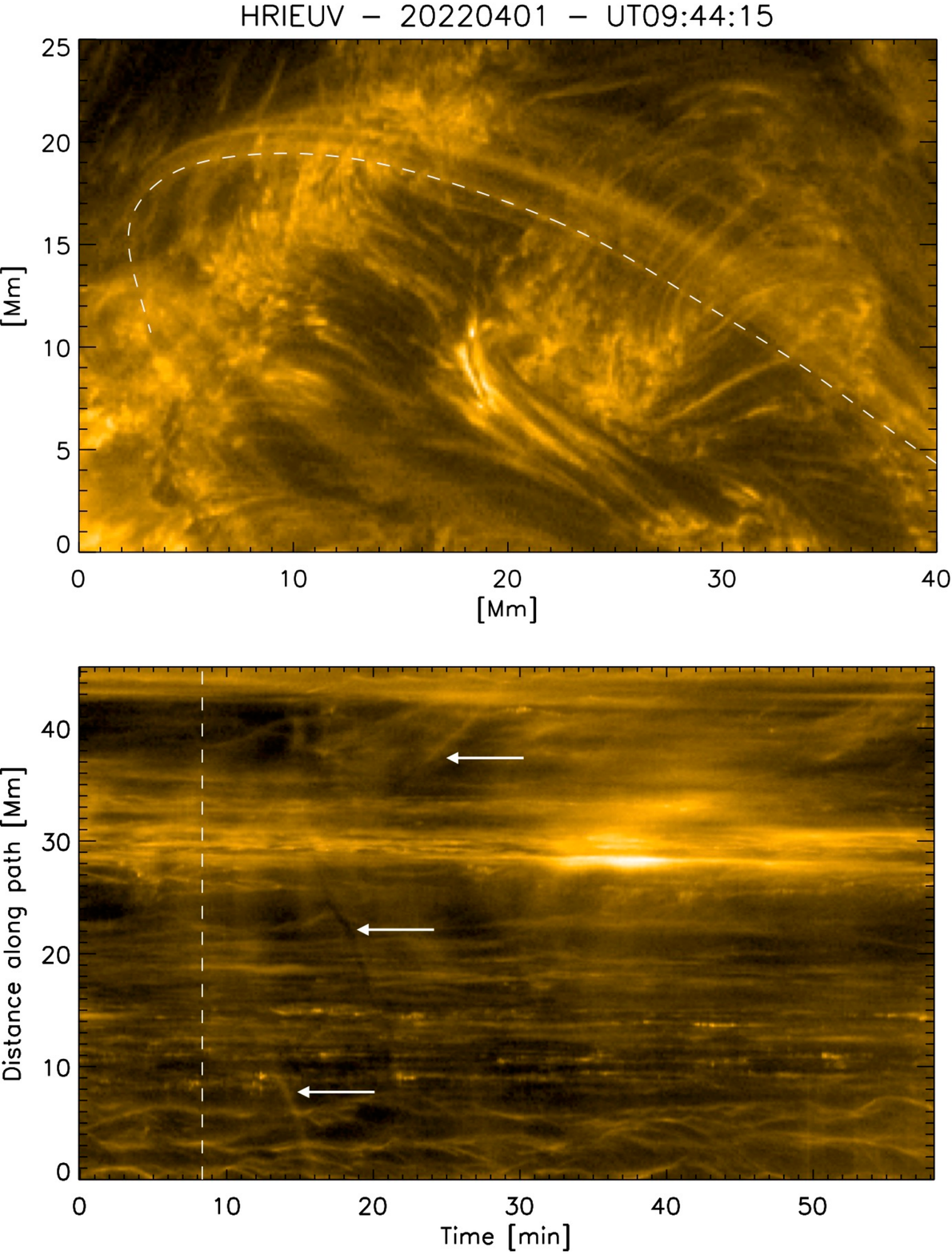}
\caption{Top: A sub-field of the 2022 April 1 dataset observed by \hrieuv (located on the right of Fig.\,\ref{fig:fibril1}). A loop bundle rooted on moss is seen composed of various strands. The strands appear first near the apex and exhibit bright and dark flows with dynamics characteristic of coronal rain. Bottom: Time-distance plot along one of the observed strands (dashed white curve on top panel). Note the fuzzy brightening events along the middle of the strand followed by bright or dark flows toward either footpoint of the strand (indicated by the white arrows). The time of the snapshot on the top panel corresponds to the vertical white dashed line. See Sect. \ref{sec:strands.and.rain}.} \label{fig:solostrand}
\end{figure}

%Patrick
In AR NOAA 2975 observed on 2022 April 1, coronal strands rooted in moss are observed to appear and disappear on a short time-scale of tens of minutes. Contrary to usual coronal strands, these appear first near the loop apex, and are seen to extend dynamically towards the footpoints in a flow-like manner. This is followed by localised dark or bright features at the loop apex with the appearance and dynamics of coronal rain (see Fig.\,\ref{fig:solostrand}). The entire event strongly resembles the 2.5D MHD numerical modelling of coronal rain by \cite{Antolin_2022ApJ...926L..29A}.

\subsubsection{M2 flare: 2022 March 2}  %%%%%%%
\label{sec:M2.flare}

\begin{figure}[ht]
\centering
\includegraphics[width=0.48\textwidth]{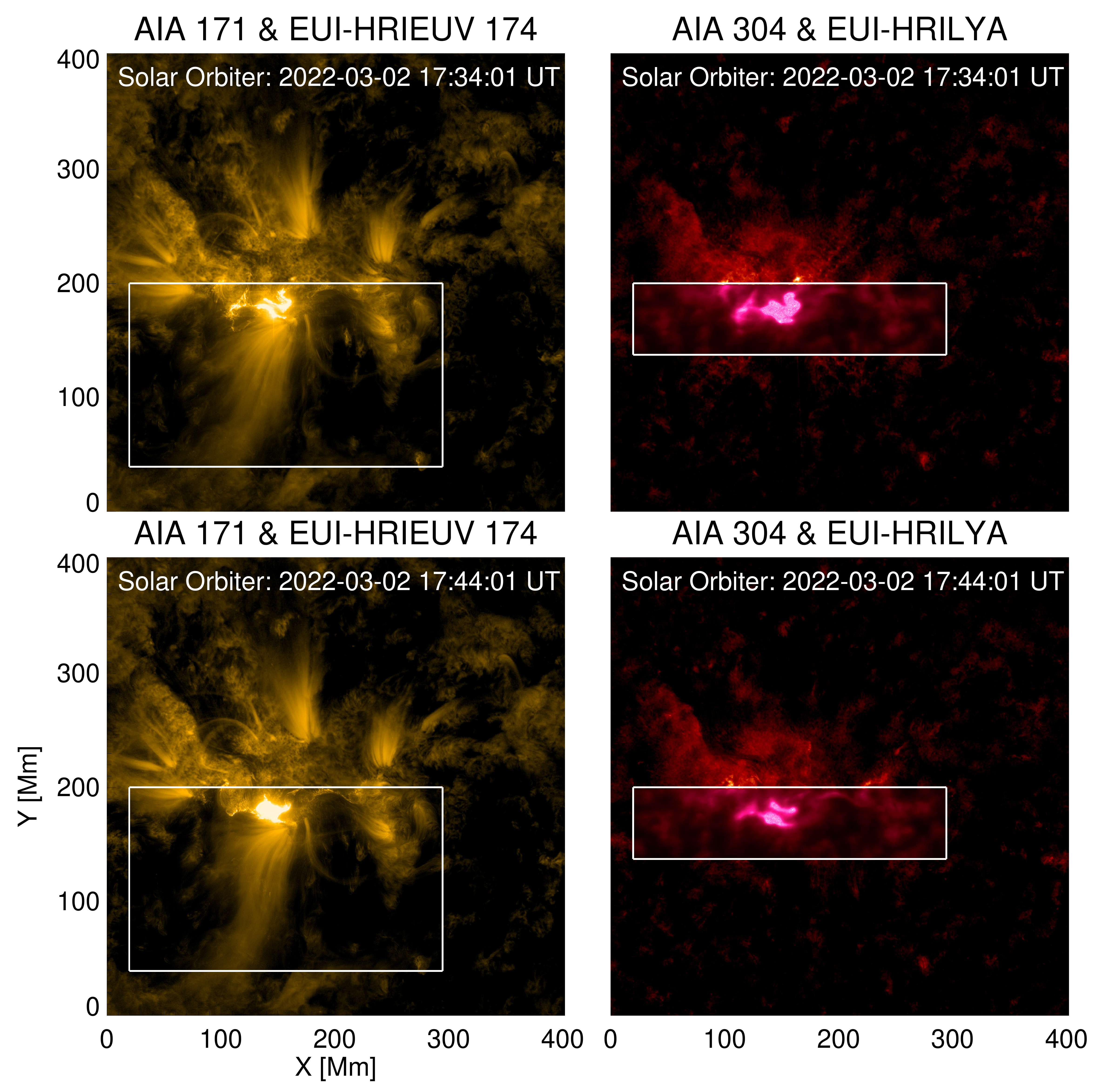}
\caption{M2 flare on 2022 March 2. Left hand side panels: combination of AIA 17.1\,nm images and \hrieuv 17.4\,nm, the latter being shown within the white box (only few seconds apart). Right hand side panels: Same, but for the combination of AIA 30.4\,nm  and \hrilya images. The bottom row is taken 10\,min later than the top row images. The white boxes delineate the boundaries of the areas showing data from EUI-HRIs. See Sect. \ref{sec:M2.flare}.} \label{fig:m2_flare-AIA_EUI-HRIs}
\end{figure}

\begin{figure}[ht]
\centering
\includegraphics[width=0.48\textwidth]{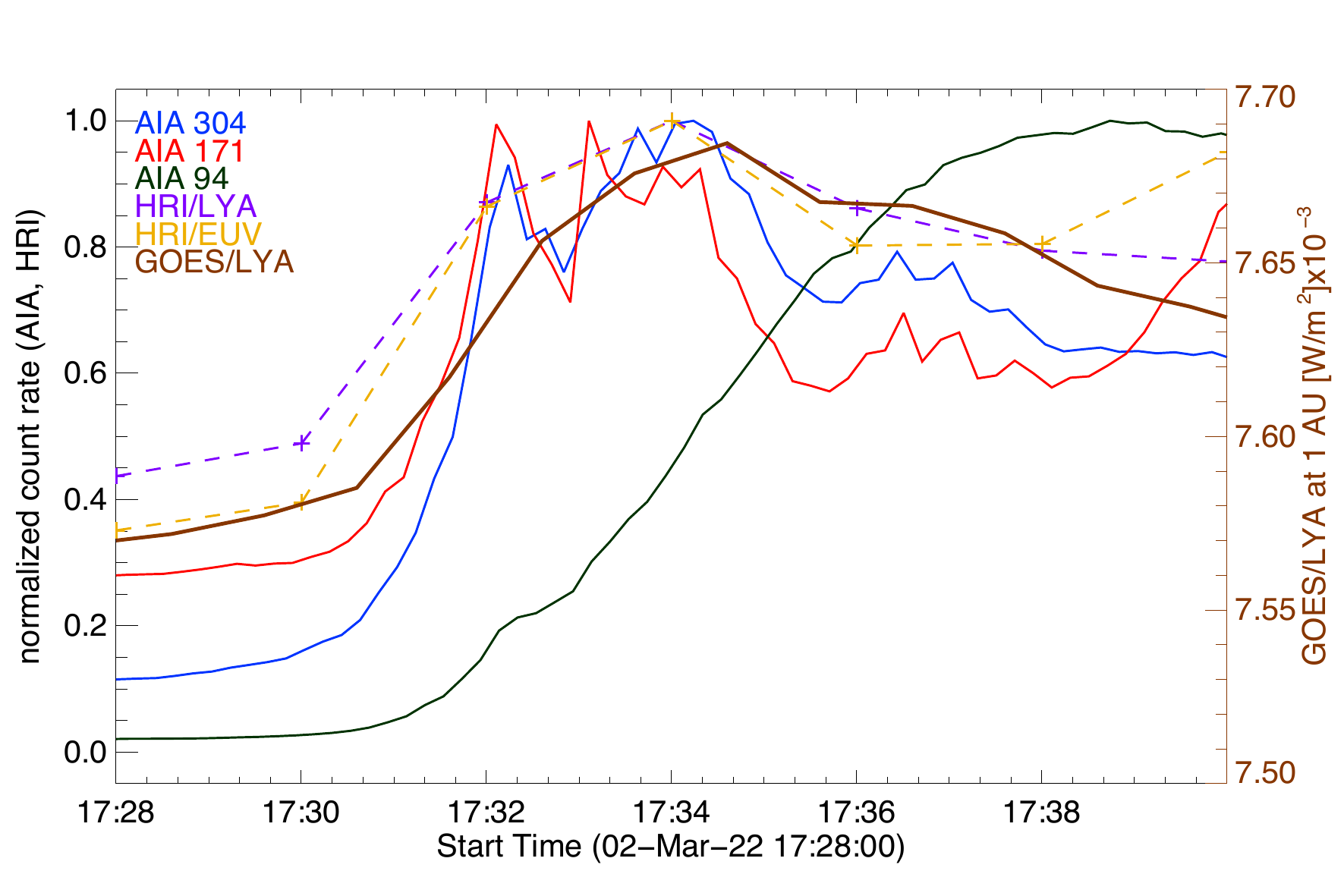}
\caption{Temporal evolution of intensity of the M2 flare on 2022 March 2. The colour-coded lines present average intensities at the flare region observed with \hrieuv, \hrilya, SDO/AIA and GOES. See Sect. \ref{sec:M2.flare}.} \label{fig:m2_flare_int_curve}
\end{figure}

%%%%%%Start text by KB mixed with Ziwen contribution
The chances to observe flares in the small FOVs of the HRIs, which are only operated a small fraction of an orbit, are not large.
Nevertheless, on 2022 March 2, during the mosaic pattern of \hyperref[Slow-Wind-Connection]{L\_SMALL\_HRES\_HCAD\_Slow-Wind-Connection}, an M2 flare was observed in active region NOAA 12958 by \hrieuv and \hrilya (Fig.\,\ref{fig:m2_flare-AIA_EUI-HRIs}). This is the largest flare seen so far in the HRI subfields.
Since the HRIs FOV only  covered the lower part of this active region, we have in Fig.\,\ref{fig:m2_flare-AIA_EUI-HRIs}  also used AIA 17.1\,nm and 30.4\,nm images to show the context of the evolution of the entire active region.

Solar Orbiter was at 0.55\,au from the Sun, meaning that the spatial resolution of the \hrieuv images (two pixel size) is 397\,km. The imaging cadence was unfortunately only 2\,min. Fig.\,\ref{fig:m2_flare-AIA_EUI-HRIs} shows the two ribbons shortly before the flare peaked at 17:34 and \hrieuv saturated with a front filter diffraction pattern.
The \hrieuv images show several small-scale brightenings during the pre-flare time.  Views from \hrieuv  and \hrilya are shown within the white boxes in Fig.\,\ref{fig:m2_flare-AIA_EUI-HRIs}. Thanks to the high resolution of \hrieuv  (left hand side panels), we can observe in detail the brightening structures in this active region, such as the double J-shaped brightening in the core region. In the right hand side panels, the lower resolution and the saturated signal of the \hrilya images only allow distinguishing the outline of the brightening structures. After the peak time, some brightenings can be found at the upper edge of the HRIs FOV appearing at the source region and propagating from west to east, forming a long thin bright band of several tens Mm, as shown in the right hand side panels. Flare loops are clearly visible in the \hrieuv image, while the flare ribbons (and flare-driven rain) are visible as two parallel bright structures in the \hrilya image.

Fig.\,\ref{fig:m2_flare_int_curve} shows that the temporal variation of the intensities observed by \hrieuv and \hrilya are in qualitative agreement with comparable observations by SDO/AIA and GOES/LYA. The \hrieuv intensity corresponds well with the SDO/AIA 17.1\,nm and 30.4\,nm, only the emission observed in SDO/AIA 9.4\,nm occurs few minutes after the rest of the presented lines. The \hrilya and GOES/LYA intensity curves corresponds well too.
% was it considered: saturation of HRI data?
% All curves are normalized to the peak value (although GOES/LYA seems not!)
%citations?

\subsection{Quiet Sun features}      %%%%%%%%%%%%%%%%%%%%%%%%%%%%%%%%%%%%%%%%%%%%%%%%%%%%
\label{QuietSun}

In the current phase of the solar cycle, most active regions appear beyond 15$^{\circ}$ solar latitude North or South, meaning that whenever Solar Orbiter was not off pointed away from disk center, the HRIs had the quiet Sun  in the FOV. Below we present a small sample of quiet Sun features and events that illustrate the quality of the obtained data.

\begin{figure}[ht]
\centering
\includegraphics[width=0.49\textwidth]{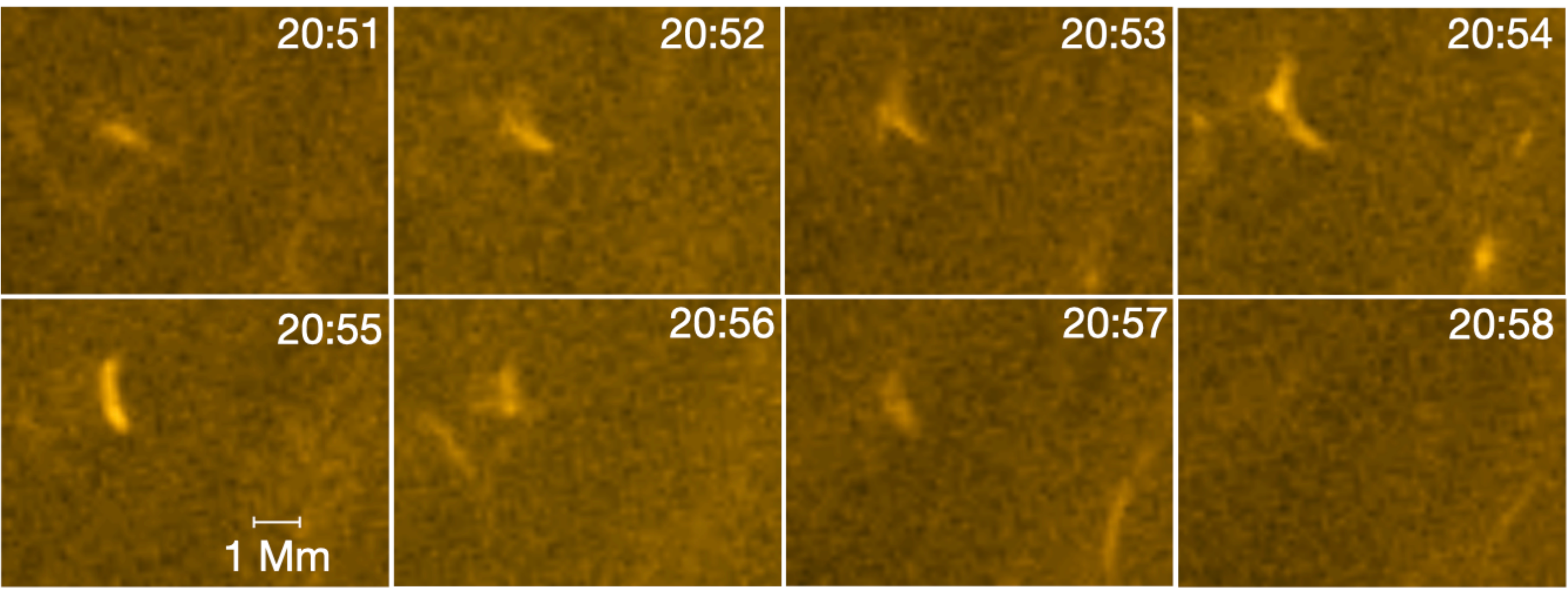}
\caption{Examples of a group of campfires observed near perihelion on 2022 March 27.  See Sect. \ref{sec:campfires}.}
\label{fig:Campfires.pdf}
\end{figure}

\begin{figure}[ht]
\centering
\includegraphics[width=0.48\textwidth]{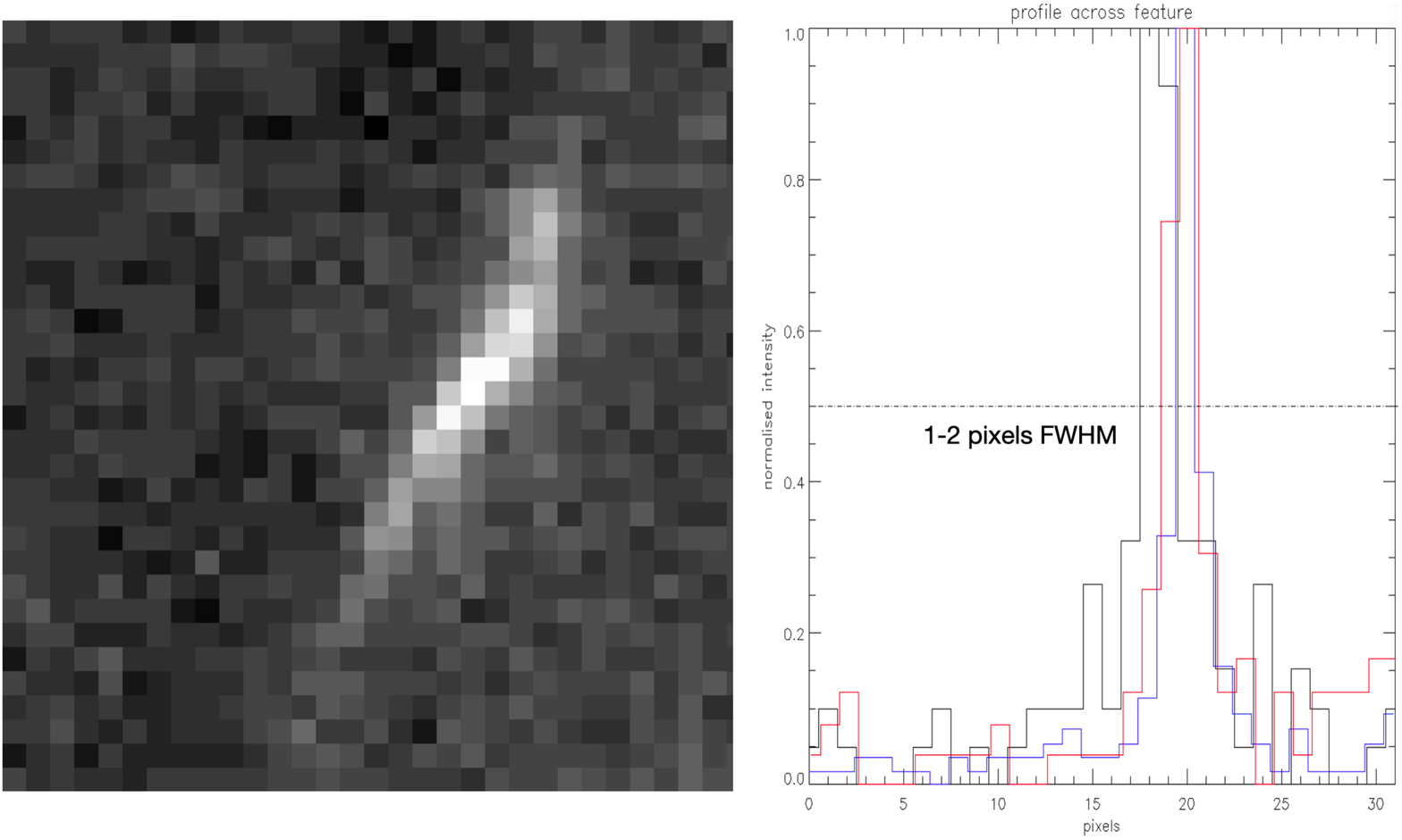}
\caption{A long and slender campfire observed on 2022 March 27. The left image shows the calibrated images (L2) with the original sensor pixelisation, each pixel corresponds to $(115~\mbox{km})^2$) on the sun.  On the right, we show various cross-cuts through the campfire demonstrating that the FWHM of the feature is 1-2 pixels.  See Sect. \ref{sec:campfires}.} \label{slender_fig}
\end{figure}

\subsubsection{Small-scale EUV brightenings\label{sec:campfires}}

On 2022 March 26, Solar Orbiter reached its first perihelion during the Nominal  Mission Phase at a distance of 0.323\,au from the Sun. The
\hrieuv observations closest to perihelion were taken on March 27, starting at 19:40 (distance 0.324\,au) and were part of the \hyperref[Coronal-Dynamics]{L\_FULL\_HRES\_HCAD\_Coronal-Dynamics} SOOP.
On this day, \hrieuv had a pixel footprint on the sun of $(115~\mbox{km})^2$. % (115 km)^2

Small EUV brightenings observed by \hrieuv, a.k.a. campfires, were first identified by \citep{2021A&A...656L...4B} in data taken at 0.556\,au. The observed campfires were typically elongated structures from 0.2 Mm to 4 Mm with aspect ratios between 1 and 5.
In Fig.\,\ref{fig:Campfires.pdf} we show a subfield, particularly rich in campfires, taken on 2022 March 27 at a distance of 0.324\,au from the Sun.  The 2022 March 27 \hyperref[Coronal-Dynamics]{L\_FULL\_HRES\_HCAD\_Coronal-Dynamics} dataset has a cadence of 1 min but several of the \hyperref[Nanoflares]{R\_BOTH\_HRES\_HCAD\_Nanoflares} data sets have cadences down to 3\,s. The visually identified sample of campfires in  Fig.\,\ref{fig:Campfires.pdf} appear somewhat smaller (none is larger than 2 Mm) but this needs to be confirmed by objective algorithmic detection.

Fig.\,\ref{slender_fig} shows an example of particular long and slender campfire demonstrating  that loop-like features are present in the quiet Sun with a width of the order of 200\,km. It also demonstrates that the spatial resolution of \hrieuv is pixel-limited.

\begin{figure}[ht]
\centering
\includegraphics[width=0.49\textwidth]{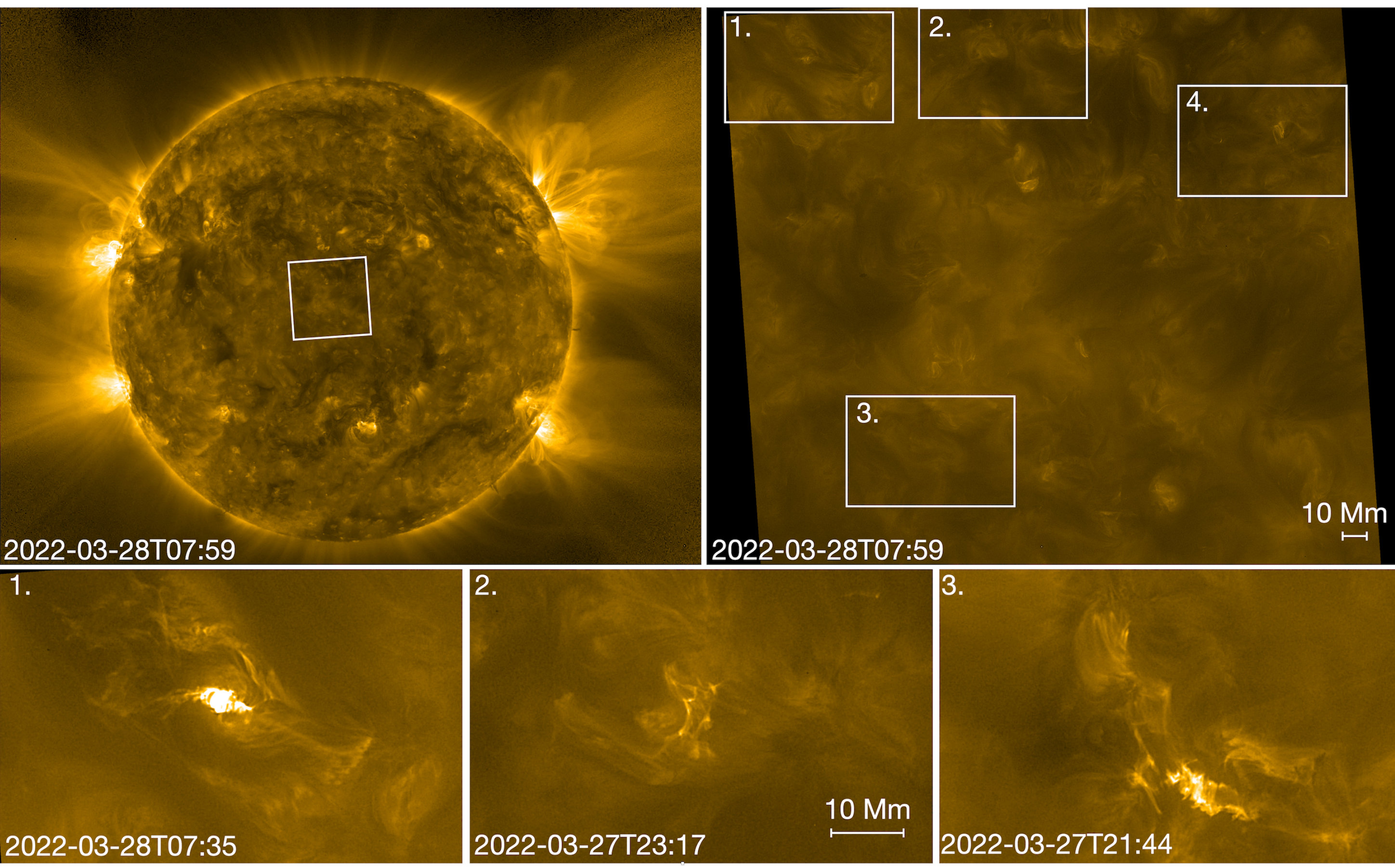}
\caption{EUV network flares observed in the quiet Sun. The top-left panel shows an FSI 17.4\,nm image with the white rectangle indicating the \hrieuv FOV that is zoomed in on the top-right panel. On this \hrieuv image, taken at a time without network flares, the location of four network flares is indicated. Three of these network flares are shown in the bottom row of the figure. The 4$^{\mbox{th}}$ event is shown in Fig.\,\ref{fig:NetworkfFlare4.pdf}. The \hrieuv pixels correspond to
$(115~\mbox{km})^2$ on the Sun. See Sect. \ref{sec:network.flares}.} \label{fig:Networkflares123.pdf}
\end{figure}

\begin{figure}[ht]
\centering
\includegraphics[width=0.48\textwidth]{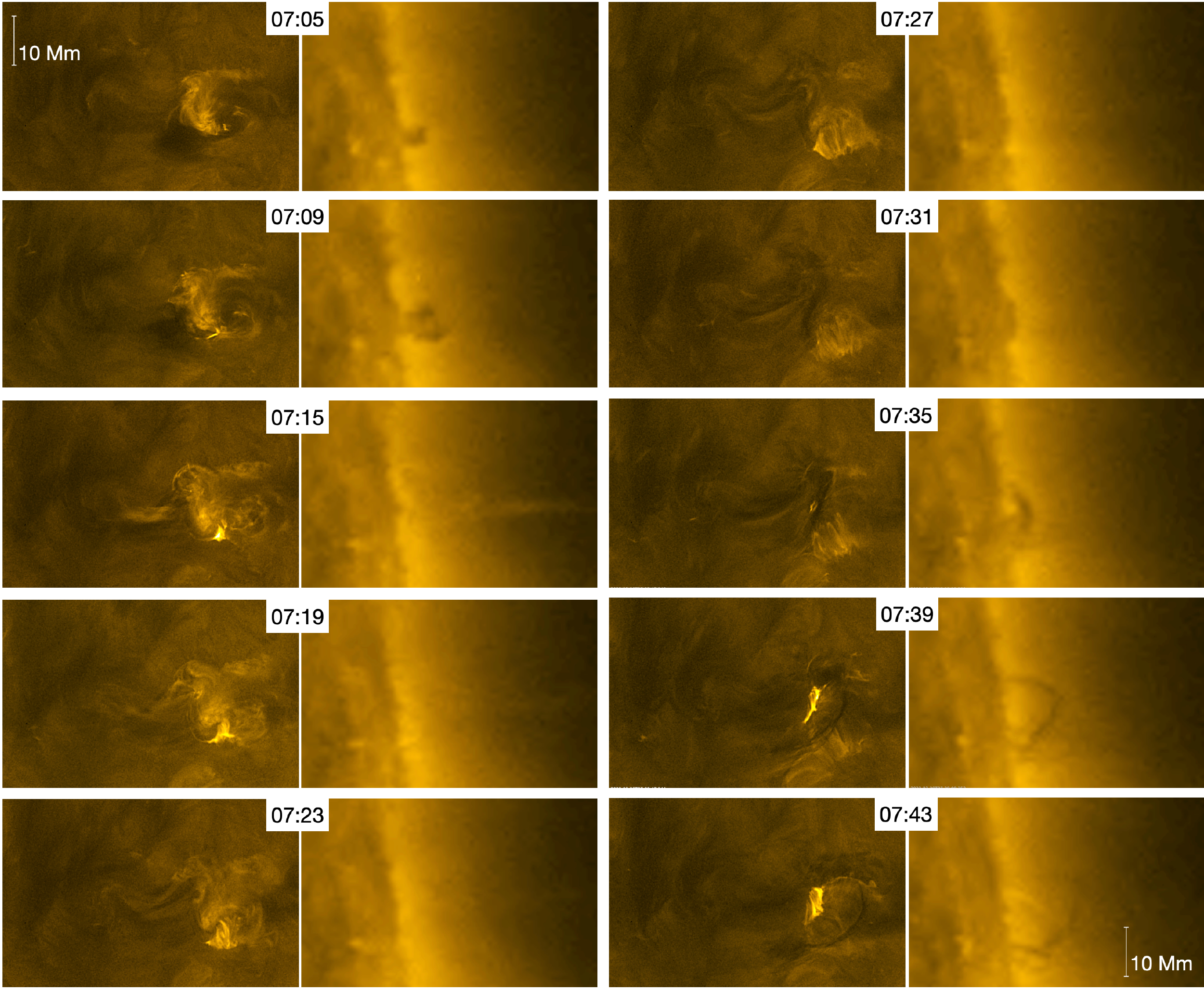}
\caption{Time evolution of an EUV network flare seen by \hrieuv on 2022 March 28 at 17.4\,nm (left) near disc center and by SDO/AIA 17.1\,nm near the limb. This event corresponds to location 4 in the top-right panel of Fig.\,\ref{fig:Networkflares123.pdf}. The off limb data confirm the eruptive character of this EUV network flare. See Sect. \ref{sec:network.flares}.} \label{fig:NetworkfFlare4.pdf}
\end{figure}

\subsubsection{EUV network flares\label{sec:network.flares}}

Also at larger scales than campfires (say $\gtrsim$ 10 Mm), flare-like brightenings are frequently seen in the quiet Sun. Fig.\,\ref{fig:Networkflares123.pdf} shows (top-right) the location of 4 such flare-like brightenings in a typical quiet Sun scenery observed by \hrieuv on March 27/28. The time evolution of the 4$^{\mbox{th}}$ event is shown in Fig.\,\ref{fig:NetworkfFlare4.pdf} with corresponding SDO/AIA 17.1\,nm imagery in quadrature showing the off-limb evolution. In X-rays, such events have been called Network Flares \cite[][]{1997ApJ...488..499K,2016A&A...596A..15A}.  The \hrieuv extreme high resolution EUV images of the quiet Sun confirm that these brightenings do indeed show many of the usual flare attributes such as pre-flare sigmoids, dimmings, ribbons and post-flare loops. At least for some of these (like in Fig.\,\ref{fig:NetworkfFlare4.pdf}) we can also confirm jets and filament eruptions, making them candidate sites for mini-CMEs \cite[][]{2009A&A...495..319I,2015Natur.523..437S}.

% Pradeep suggested to cite also the Alphonse Sterling and Moore work on mini-filament eruptions

%    \item surprisingly large flare for quiet sun
%    \item observed near the limb in AIA 171, 303 and 131
%    \item looks very 'classical', ie sigmoid filament, flare ribbons, post flare loops
% https://chat.observatory.be/eui/pl/9csto3nxqjy9zy99dgmaj59f9a

\begin{figure}[ht]
\centering
\includegraphics[width=0.2\textwidth]{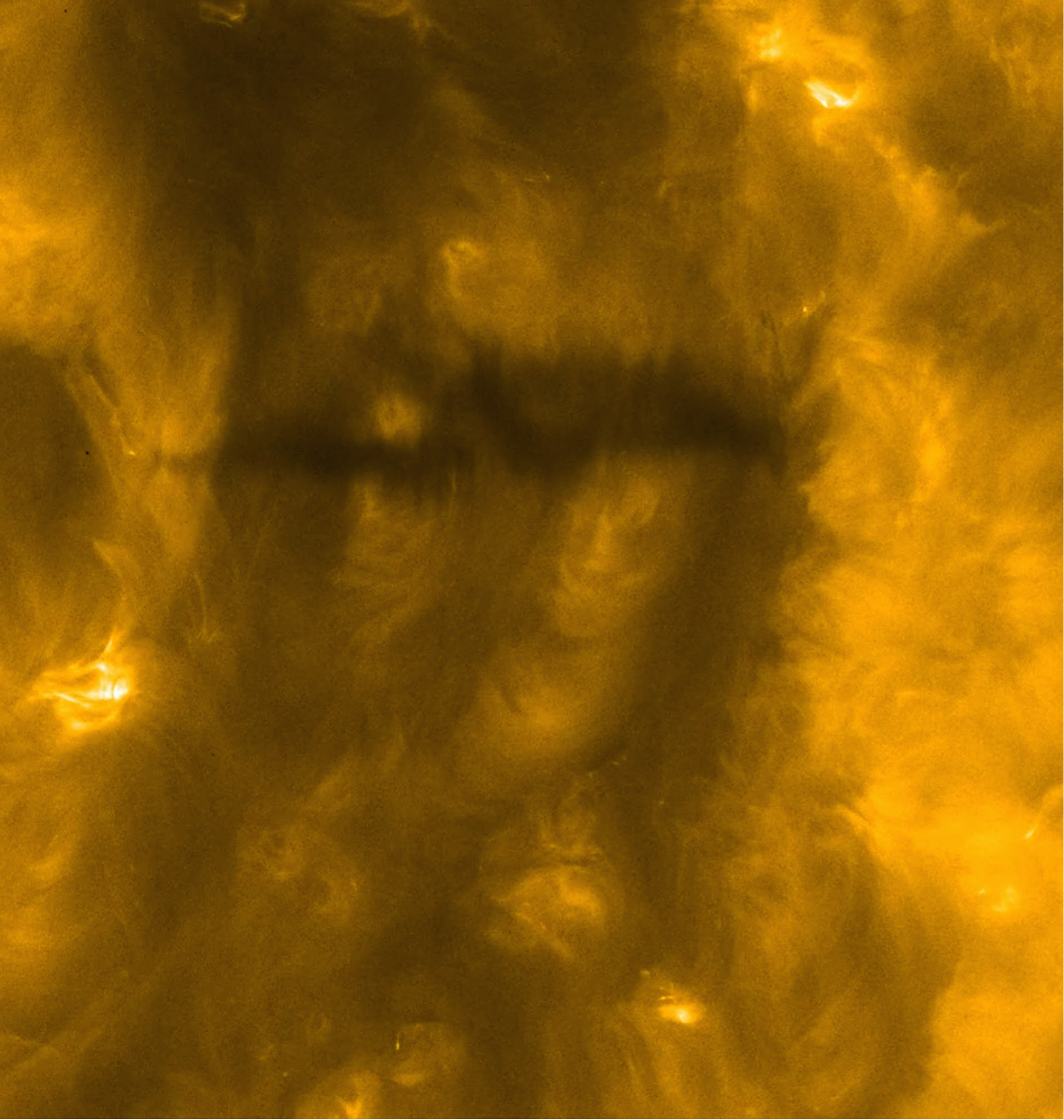}
\includegraphics[width=0.2\textwidth]{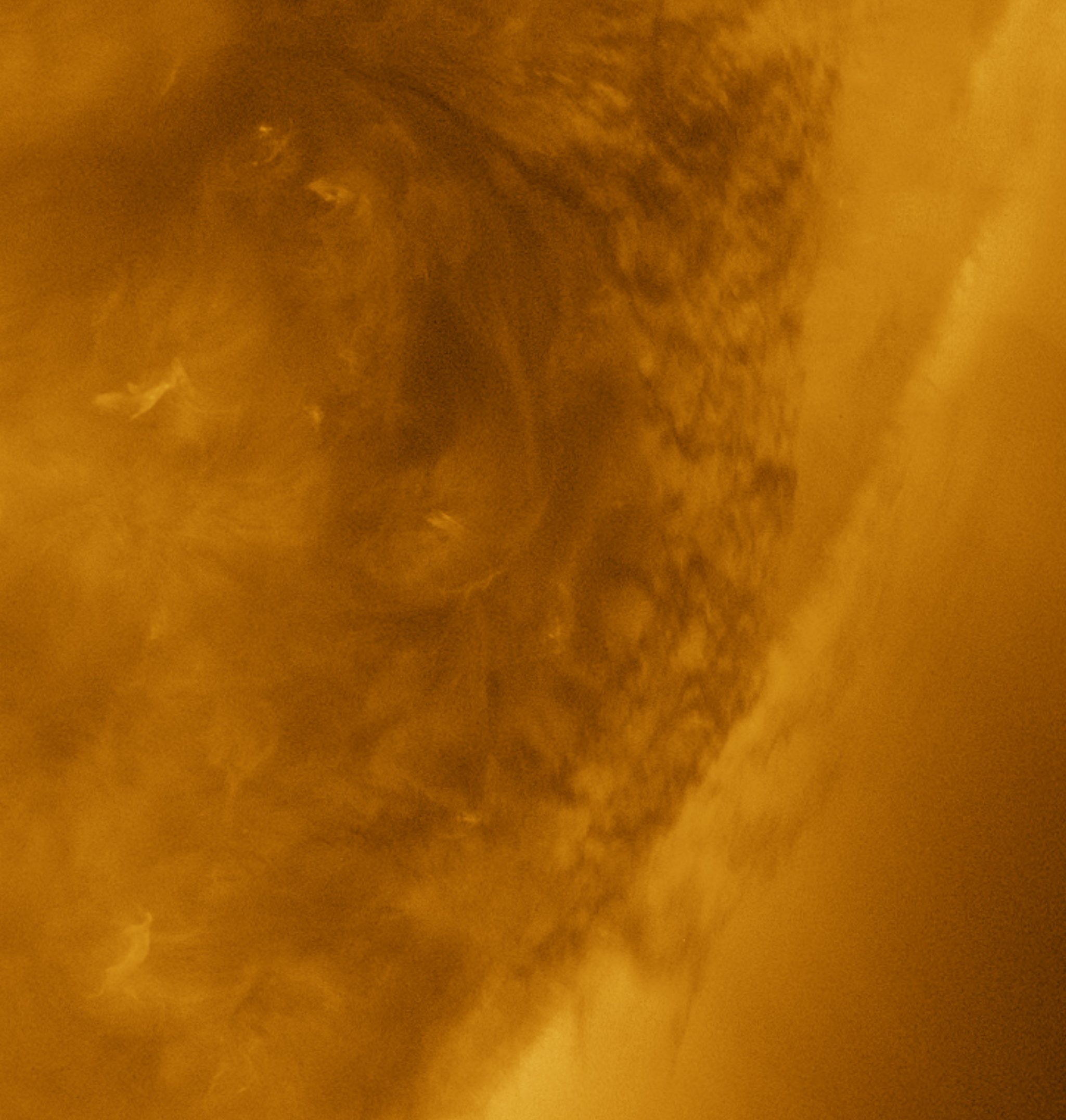}\\
\includegraphics[width=0.2\textwidth]{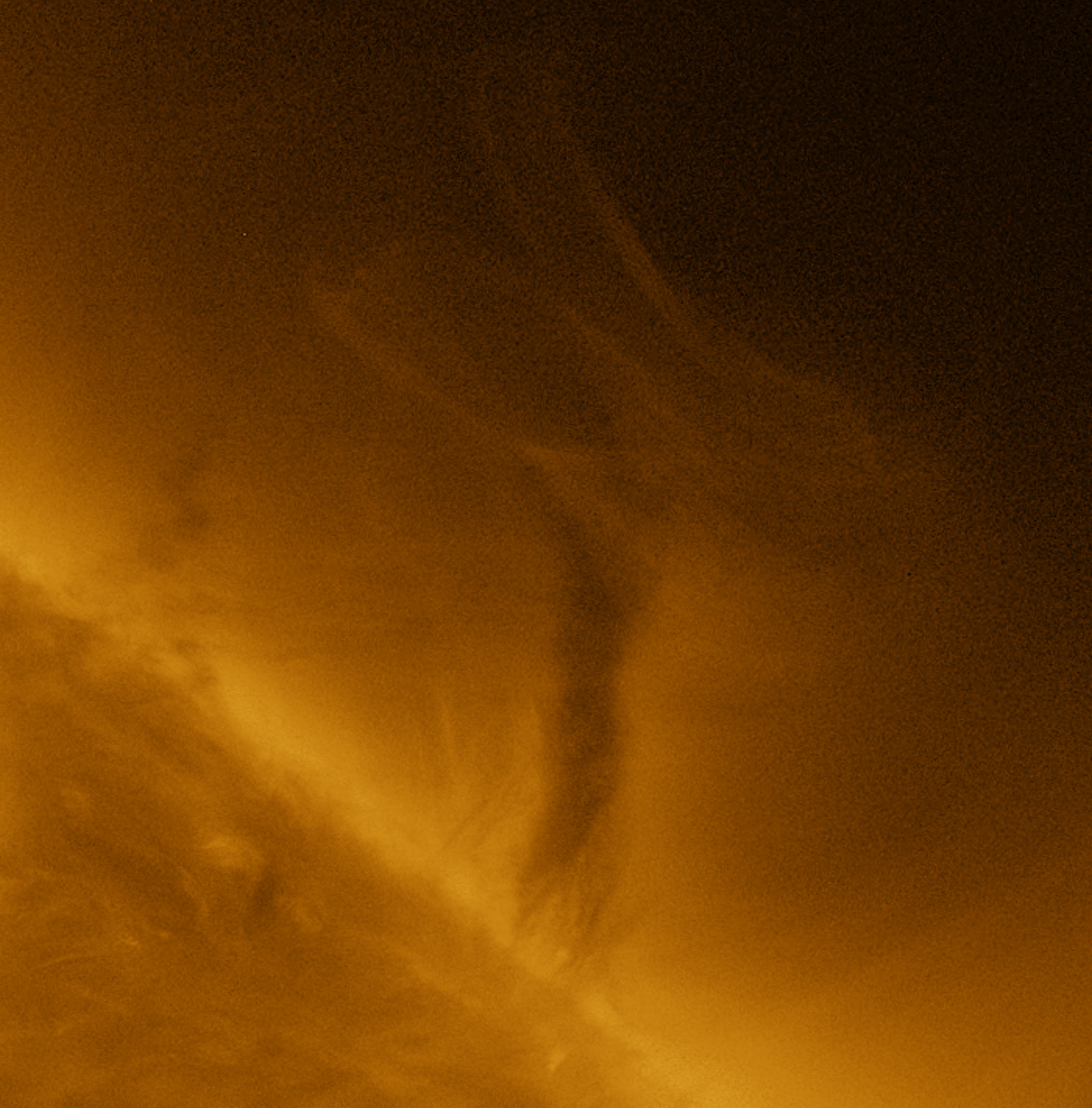}
\includegraphics[width=0.2\textwidth]{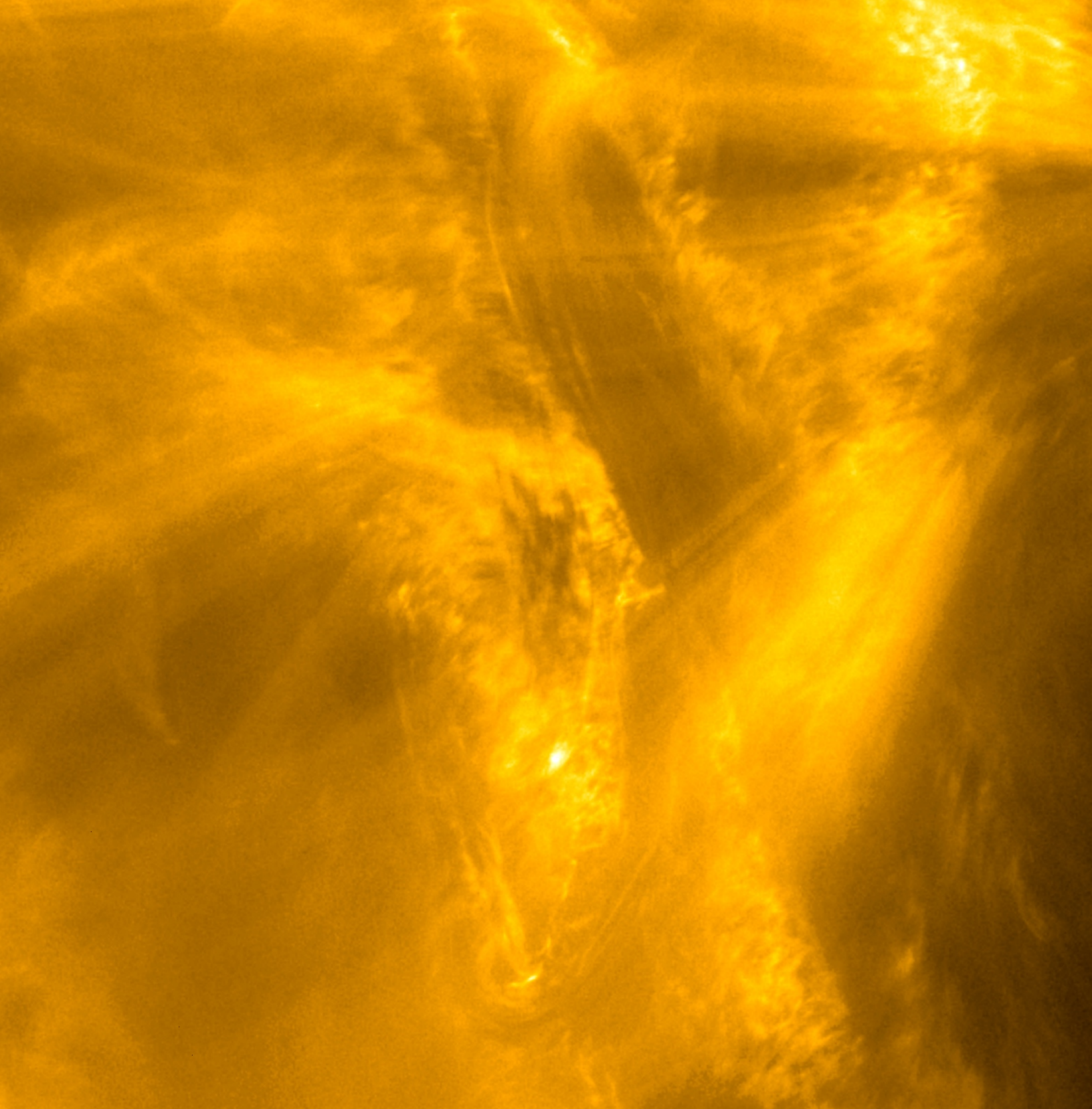}\\
\caption{Examples of filaments and prominences observed in \hrieuv in 2022 March. See Sect. \ref{sec:filaments}.} \label{fig:fil_20220318}
\end{figure}

\begin{figure}[ht]
\centering
\includegraphics[width=0.5\textwidth]{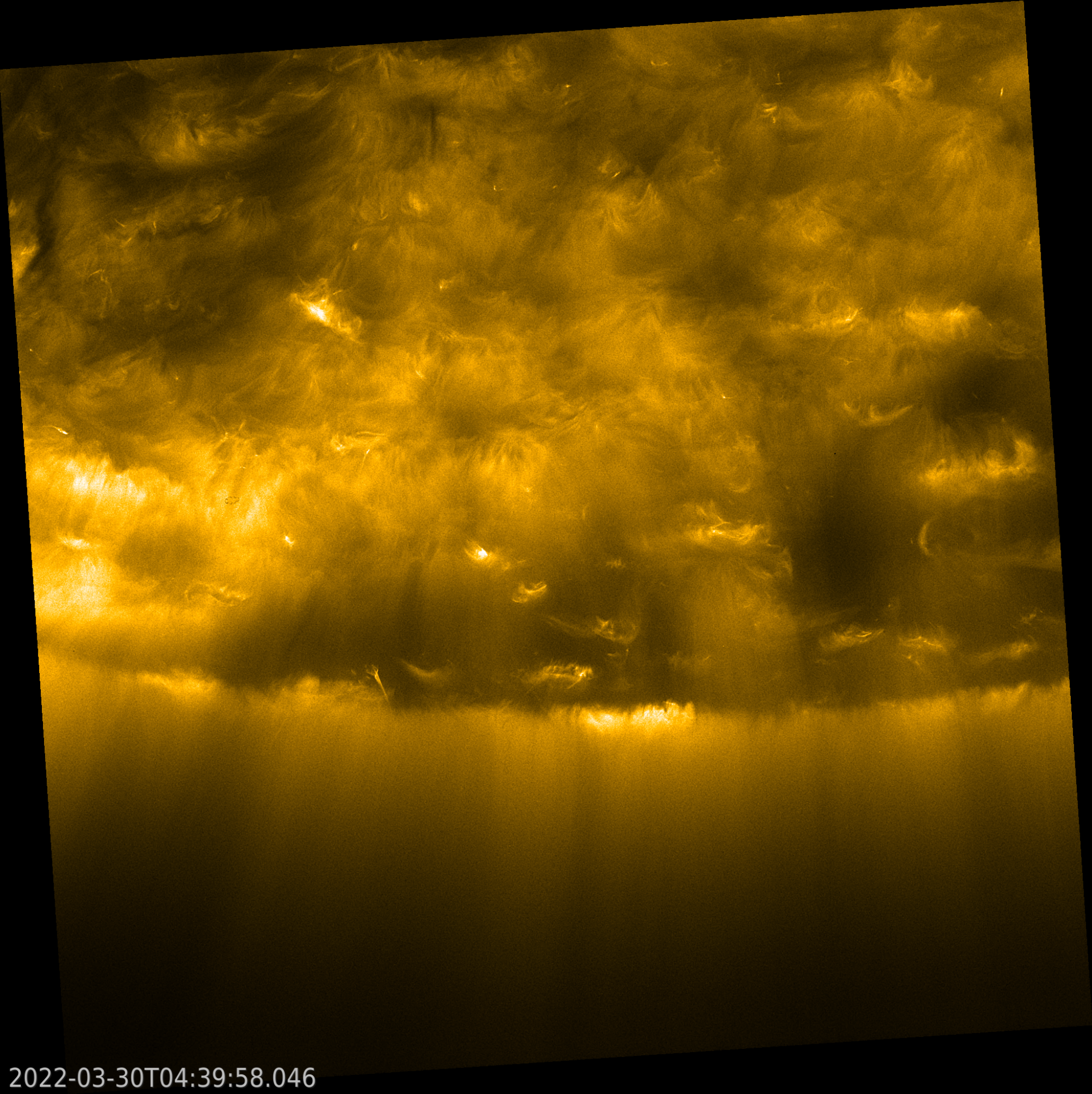}
\caption{South polar coronal hole imaged by \hrieuv on 2022 March 30. Solar north is up, west is to the right. See Sect. \ref{sec:polar}.} \label{fig:polar}
\end{figure}

\subsubsection{Polar coronal holes\label{sec:polar}}

Fine-scale structure and dynamics of coronal holes is of particular interest for studies of the fast solar wind origin \citep[e.g. ][]{2007Sci...318.1580C, 2015LRSP...12....7P}. Due to the progression of the ascending phase of the solar cycle, polar coronal holes were shrinking in 2022, so it was important to observe these structures at high spatio-temporal resolution early in the Solar Orbiter mission. During the first perihelion passage in 2022, this was done on three occasions: on March~6, March~30, and April~4--5. The R\_SMALL\_HRES\_MCAD\_Polar-Observations SOOP was used (see Table~\ref{datasetTable}).

On March 30 Solar Orbiter was situated at the distance of 0.33~au from the Sun, and \hrieuv reached the two-pixel spatial resolution of around 240~km. This was the highest ever spatial resolution reached in coronal hole observations. An excellent cadence of 3~s allowed observing numerous dynamic fine-scale structures in the south polar coronal hole (see Figure~\ref{fig:polar}), including ``bright points'', plumes, plumelets, jetlets, and jets (Chitta et al. 2022, submitted). Numerous campfires were visible in the adjacent quiet Sun area.

Similar datasets were taken for the north polar coronal hole on March~6 and April~4--5,  when the distance from the spacecraft to the Sun was 0.5~au and 0.37~au respectively (Table~\ref{datasetTable}). The \hrieuv spatial resolution of around 270~km was reached on April~4. Even if 7 \hrieuv images were taken at the very high cadence of 2~s on March 6, the typical cadence was 30~s in both datasets.

\subsubsection{Filaments and prominences observations\label{sec:filaments}}

The first perihelion passage of Solar Orbiter has also permitted EUI to take close up observations of filaments and prominences at high cadence and spatial resolution. On the disk, the  width of their core fine structure  has been resolved by \hrieuv down to the limit of the H$\alpha$ instrument resolution ($\approx 0.2\arcsec$). In H$\alpha$  the threads have a width distribution centered at 0.3\arcsec, which means $\approx$ 225\,km on the
 Sun \cite[for a review of the observations see][]{Parenti2014}.
Fig.\,\ref{fig:fil_20220318} shows some example of filaments seen mostly in absorption by \hrieuv. Similar absorption features are also observed by \hrieuv in coronal rain events (see section~\ref{sec:rain}). The top left panel shows a filament which was followed for about half of an hour on 2022 March 18 at 10:30UT at a cadence of 5\,s, allowing to detect fast intensity variation at small scales along and across the structure. Full Sun images show local activity  on 2022 March 17, which led to the formation of the filament and the opening of a small coronal hole. The merging of the polar coronal hole and the dimming region formed by the eruption of the filament is discussed in Ngampoopun et al. (2022; in preparation, this issue).

The second panel on the top row of Fig.\,\ref{fig:fil_20220318} shows a filament close to the limb, that was observed during the Full Disc Mosaic campaign on 2022 March 7 (see Section 2.2). Both the on disk and off disk part of the filament shows a complex fine structure, highlighted by dark and bright alternating wavy shaped features.  During the same campaign, we also observed the prominence shown on the bottom left. This has a tornado-like shape, with thin bright and dark threads and a bright extension at the base of the coronal cavity.  These observations are very promising for future quiescent prominence and filament studies, as they provide elements to derive the fine scale morphology, and characterize the dynamics from possible injected plasma and waves activity. The bottom-right panel shows part of an AR filament observed in 2022 March 30. During the 30\,min \hrieuv high cadence sequence, the filament  was quite active, with  brightenings in threads and fine dark features levitating on top of the main dark body. The cadence of 3 -- 5\,s that was chosen for all the sequences shown in Fig.\,\ref{fig:fil_20220318} appear to be adequate for studying the dynamics at such a small scales.

\subsection{Eruptions}      %%%%%%%%%%%%%%%%%%%%%%%%%%%%%%%%%%%%%%%%%%%%%%%%%%%%
\label{Eruptions}

\begin{figure}[ht]
\centering
\includegraphics[width=0.49\textwidth]{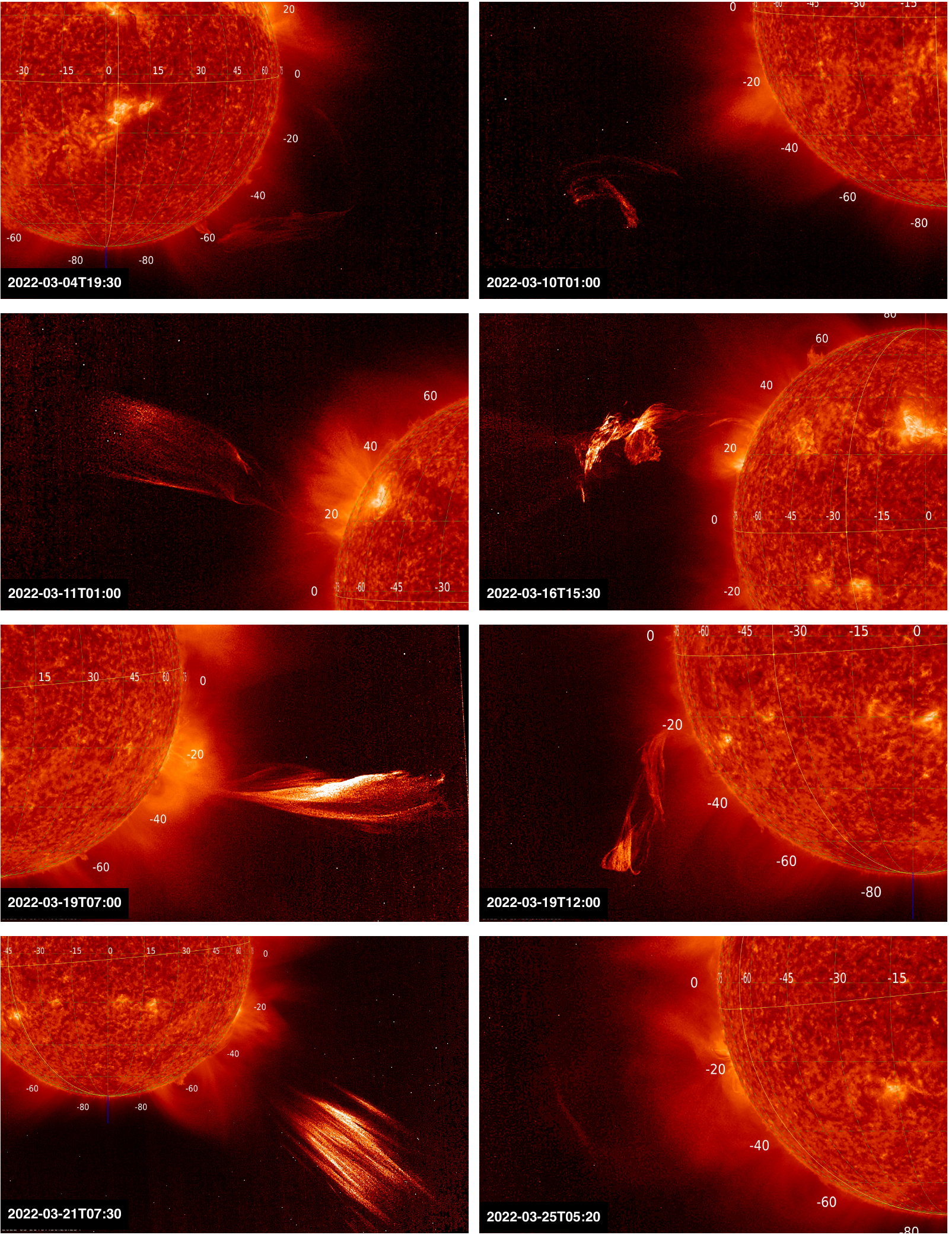}
\includegraphics[width=0.49\textwidth]{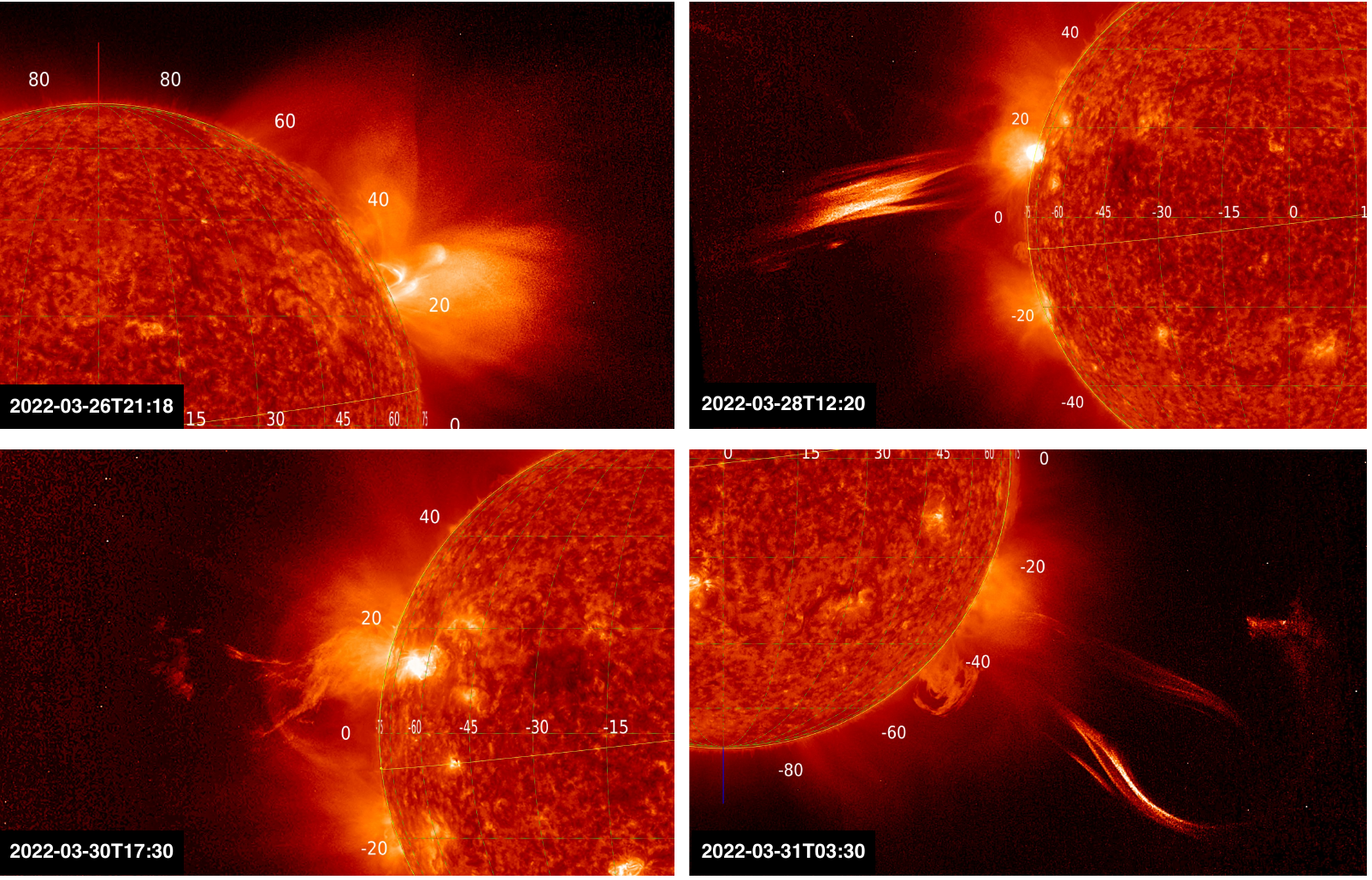}
\caption{Mosaic of prominence eruptions observed by FSI in the 30.4\,nm passband in 2022 March. The 'enhance off limb' functionality of JHelioviewer \citep{Mueller2017} was used when creating these graphics. See Sect. \ref{Eruptions} and \ref{sec.east.limb.2022.March.30}.} \label{fig:mosaiceruptions1}
\end{figure}

Due to varying distance to the Sun, the FSI FOV (from disc center to edge) changed from 4 $R_\odot$ on 2022 March 2, to 2.3 $R_\odot$ at perihelion and back to 2.8 $R_\odot$ on 2022 April 6. This large FOV,  unprecedented for an EUV imaging telescope, allowed us to monitor the early evolution of eruptions.  Table~\ref{T-ListEruptions} lists the approximate starting time, the shape and the greatest height reached by the eruptions observed by FSI during the period. Some example prominence eruptions are shown in Fig.\,\ref{fig:mosaiceruptions1}. They appear in a multitude of shapes: surge-like, loop-like, curled-like eruptions and have different kinematic behavior, from slow rising to fast eruptions. In the following subsections, we highlight a number of eruptions with particularly good coverage by other instruments or with space weather relevance.

\begin{table*}[!ht]
\renewcommand{\arraystretch}{1.2}
\begin{center}
\caption{Eruptions observed by FSI between 2022 March 2 and 2022 April 6.}
\label{T-ListEruptions}
\begin{tabular}{cccc}
\hline\hline
Start Date	&	Position	&  FSI channel	&	Comments	\\

\hline
2022 March 04	&	SW and E  &  304	 &	two prominences at SW (18:00, loop-like opening, up to 1.65~$R_\odot$) \\
            &             &          &   and E (21:00, jet-like, up to 2.25~$R_\odot$)	\\
2022 March 05	&	SE  &  304, 174 &	small prominence  (12:30, along a loop, up to 1.24~$R_\odot$)	\\
2022 March 06	&	NE  &  304	 &	small prominence (03:00, loop-like, up to 1.31~$R_\odot$)	\\
2022 March 08	&	NE  &  174	 &	 NE (08:10, jet-like?, up to 2.2~$R_\odot$)	\\
2022 March 08	&	SE  &  174	 &	to the outer FOV \\
   &             &          &  (21:00, fan-like, extended concave-out, up to 3.12~$R_\odot$)	\\
2022 March 09 & SE  &  304	 &	 far in the FOV (19:30, twisted, 2.45~$R_\odot$)	\\
2022 March 10	&	NW quadrant  &  174	 &	 on-disk (18:30 dimming; 21:30 post-eruption arcade)	\\
2022 March 10	&	E  &  304	 &	 end FOV (19:00, jet-like, 3.24~$R_\odot$)	\\
2022 March 10	&	E  &  304	 &	 end FOV (23:30, fan-like, 3.15~$R_\odot$)	\\
2022 March 12	&	W  &  174	 &	 to the end FOV (06:00, elongated sinusoidal?, up to 2.85~$R_\odot$)	\\
2022 March 13	&	E  &  304, 174	 &	two small  (00:30, 05:00, loop opening, 1.50~$R_\odot$)	\\
2022 March 14	&	SW  &  174	 &	big  (17:20, loop-like opening, 2.24~$R_\odot$)	\\
2022 March 16	&	E  &  304	 &	2 curled prominences (13:00, 14:30, 2.65~$R_\odot$)	\\
2022 March16	&	SE, NE, SW  &  174	 &	elongated curved  SE (08:00, 2.60~$R_\odot$) \\
   &             &          &  2 eruptions, 14:10 NE and SW 	\\
2022 March17	&	W  &  174	 &	small  (06:30, 2.10~$R_\odot$)	\\
2022 March 18	&	W  &  174	 &	small (11:00, concave-out, 2.30~$R_\odot$)	\\
2022 March 19	&	W, SE  &  304	 &	prominence W (06:00, fan-like, 2.6~$R_\odot$) and SE (10:30, curled, 1.81~$R_\odot$)	\\
2022 March 19	&	SE  &  174	 &	 (10:00, curled, 1.8~$R_\odot$)	\\
2022 March 20	&	NE  &  304, 174	 &	prominence  (08:00, curled, 2.32~$R_\odot$)	\\
2022 March 21	&	SW  &  304, 174	 &	 (05:30~UT, fan-like, 3.25~$R_\odot$)	\\
2022 March 24	&	SE  &  174	 &	small  (11:30, loop-like, 1.50~$R_\odot$)	\\
2022 March 25	&	SE  &  304, 174	 &	 (05:00, loop-like, 1.77~$R_\odot$)	\\
2022 March 26	&	NW  &  304, 174	 &	 (19:30, loop-like, 2.27~$R_\odot$)	\\
2022 March 27	&	E  &  304, 174	 &	2 small eruptions (13:00, 19:00, loop-like, 1.5~$R_\odot$)	\\
2022 March 28	&	E  &  304, 174	 &	prominence  (11:20, loop-like + fan, 2.3~$R_\odot$) \\
            &      &     & M4 flare and halo CME arriving at Earth on 2022 March 31	\\
2022 March 30	&	NW, E  &  304, 174	 &	 at NW (05:30, loop-like, 2.07~$R_\odot$), and E (14:00, ragged, 1.9~$R_\odot$)	\\
2022 March31	&	SW  &  304, 174	 &	(02:30, loop-like, 2.30~$R_\odot$)	\\
2022 April 02	&	NE, SE  &  304, 174	 &	prominences (13:00, ragged, 2.7~$R_\odot$)	\\
2022 April 03	&	SE  &  304	 &	prominence  (15:00, untwisting to loop-like, 2.7~$R_\odot$)	\\
2022 April 04	&	SE  &  304	 &	big filament (10:30, faint elongated off-limb, ~2.11~$R_\odot$)	\\
2022 April 05	&	SW  &  174	 &	 (13:00, fan-like, 2~$R_\odot$)	\\
2022 April 06	&	SW  &  304, 174	 &	prominence  (22:00, ragged, 1.7~$R_\odot$)	\\
\hline
\end{tabular}
\end{center}
\end{table*}

%\begin{figure}[ht]
%\centering
%\includegraphics[width=0.49\textwidth]{img_march_promin_eruptions_2.pdf}
%\caption{Continuation, Mosaic of prominence eruptions observed by FSI in the %30.4\,nm passband in 2022 March. The 'enhance off limb' functionality of %JHelioviewer \citep{Mueller2017} was used when creating these graphics. See Sects. %\ref{Eruptions} and \ref{sec.east.limb.2022.March.30}.} %\label{fig:mosaiceruptions2}
%\end{figure}

\subsubsection{C2.8 flare: 2022 March 10\label{sec:flare.2022.Mar.10}}

\begin{figure}[ht]
\centering
\includegraphics[width=0.48\textwidth]{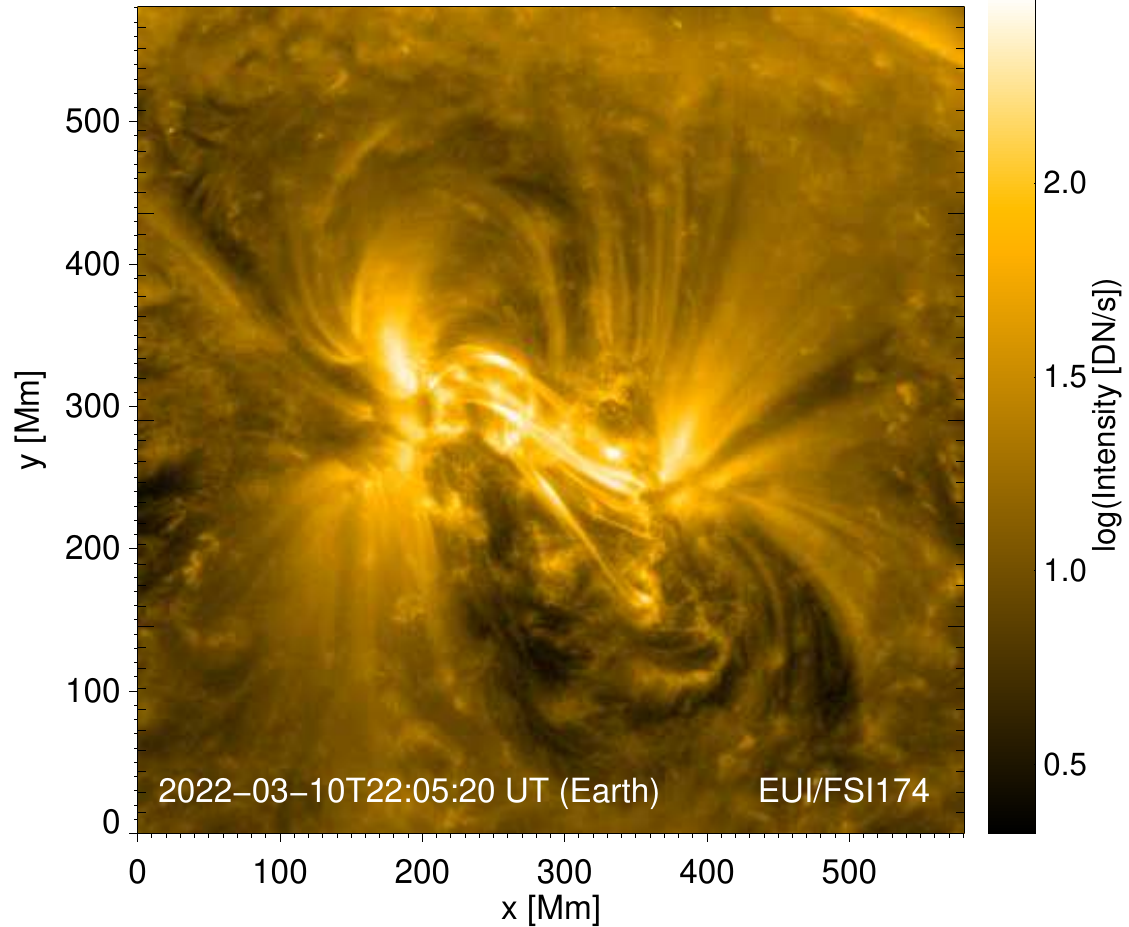}
\caption{FSI observations in the 17.4\,nm passband of the C2.8 flare on 2022 March 10. The images present the arcades of the post-flare loops. See Sect.\ref{sec:flare.2022.Mar.10}.
% (See also movie "c28\_flare.mp4")
}. \label{fig:flare_c28_fsi}
\end{figure}

The C2.8 flare from active region NOAA 12962 on 2022 March 10 (GOES X-ray peak at 20:33)  was observed by FSI as a classical two-ribbon flare, and later a post-flare arcade, with a cadence of 10\,min in the 304 channel and 30\,min in the 174 channel (see Fig.\,\ref{fig:flare_c28_fsi}).
It was also observed by STIX \citep{STIX} and EPD \citep{EPD} on Solar Orbiter. The Earth was separated by 7.8$^\circ$ from Solar Orbiter. From the Earth's perspective the flare was near the central meridian and associated with a partial halo coronal mass ejection (CME) that eventually led to a moderate geomagnetic storm (Kp = 6) on March 13 and 14. The evolution of the CME shock and its effect on ion acceleration was studied by Walker et al., 2022 (this volume, in prep).

\subsubsection{Limb CME: 2022 March 21 }
\label{section:EPDevent}

\begin{figure}%[h]
\centering
\includegraphics[width=0.48\textwidth]{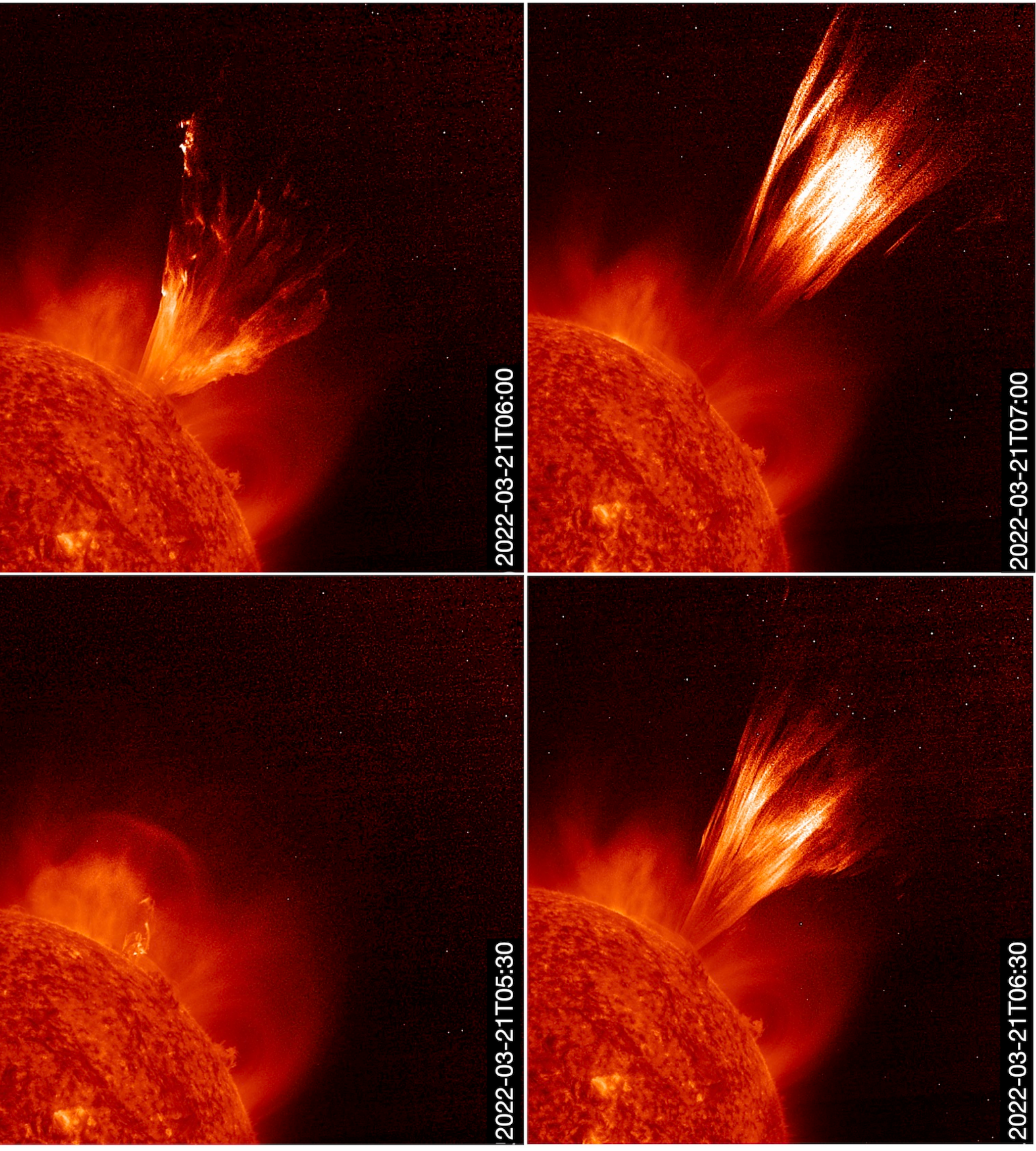}
\caption{Time sequence of the eruption on 2022 March 21, as seen by FSI in the 30.4\,nm passband. The 'enhance off limb' functionality of JHelioviewer \citep{Mueller2017} was used when creating these graphics. See Sect. \ref{section:EPDevent}.}
\label{21March}
\end{figure}

Starting around 05:30 UT on 2022 March 21, an eruption was observed by FSI at the SW limb (see Fig.\,\ref{21March}) that led to a partial halo CME observed from the Earth. The CME was associated with a Type-II radio burst \cite[measured by RPW,][]{RPW}, with X-ray emission (observed by STIX) and with a wide SEP event measured by EPD, SOHO and STEREO-A (34$^\circ$ to the east of the Earth). Solar Orbiter was at 0.34\,au from the Sun, and 44$^\circ$ west of the Sun-Earth line. The source location of the CME was located close to the west limb as seen from Solar Orbiter, at least partially occulted. The CME was fast, with speeds above 1000\,km\,s$^{-1}$.
At that time EUI was executing the Slow-Wind-Connection SOOP and observed in the 30.4\,nm passband of FSI with a cadence of 30\,min and in the 17.4\,nm passband of FSI with 10\,min cadence. A time sequence of the eruption can be seen in Fig.\,\ref{21March}.

%See Fig.\,\ref{fig:March21_EPDevent_FSI304}

%This is the old figure, in case we want it back
%\begin{figure}[h]
%\centering
%\includegraphics[width=0.4\textwidth]{Marc%h21_EPDevent_FSI304.pdf}
%\caption{Surge? eruption seen in FSI304.} %\label{fig:March21_EPDevent_FSI304}
%\end{figure}

%\color{blue} %% BEGIN 'TO  BE REMOVED'
% https://chat.observatory.be/eui/pl/76poh3zucjru3f%jpx3rcd7c8ar \newline
% see also ESA press release infographic
%\color{black} %% END 'TO  BE REMOVED'

\subsubsection{East limb eruption: 2022 March 30\label{sec.east.limb.2022.March.30}}

On 2022 March 30, at around 14:00~UT, FSI observed in the 30.4\,nm channel a prominence erupting at the East limb (see lower left panel of Fig.\,\ref{fig:mosaiceruptions1}) which was further observed by SolOHI  \citep{SolOHI}. Prominence material is still visible in the FSI FOV at around 20:30~UT. At around 17:30 a flare was observed on the disk (N15W30 Stonyhurst coordinates) and a bright loop is visible off-limb overlapping with the prominence but not disturbing its evolution. This indicates that the prominence was situated far away from the flare.
By inspecting the SDO/AIA304 and STEREO-A/EUVI304 \citep{Howard2008} movies one could see an extended filament erupting at the East of the flare.

FSI observed in the 17.4\,nm passband faint material erupting at around 14:00 at the East limb followed by a big dimming off-limb at around 17:30. Material moving out is still observed at 20:50. A large EUV wave is observed on-disk.

The eruption was associated with a flux-rope like coronal mass ejection observed by STEREO-A/COR2 and SOHO/LASCO-C2 \citep{Brueckner1995} coronagraphs at West limb at 18:23 and 18:12 respectively.

%\begin{figure}[ht]
%\centering
%\includegraphics[width=0.5\textwidth]{img_20220330_fsi304_fsi174_eruption_E.pdf}
%\caption{FSI304 and FSI174 observations of the eruption observed at E limb on March 30 2022.} \label{fig:eruption30MarchE}
%\end{figure}

%\begin{figure}[ht]
%\centering
%\includegraphics[width=0.5\textwidth]{img_20220330_fsi174_eruption_NW.pdf}
%\caption{FSI174 observations of the eruption observed at NW limb on March 30 2022.} \label{fig:eruption30MarchNW}
%\end{figure}

\subsubsection{North-east limb eruption: 2022 April 2}

On 2022 April 2, FSI observed in the 30.4\,nm channel a filament erupting at the NE limb (as seen from Solar Orbiter) between 13:00 and 13:30~UT. It was associated with an M3.9-class flare. The event was also captured by several other remote-sensing instruments on Solar Orbiter such as SPICE, STIX, and SoloHI. Interestingly, the erupting filament was also monitored a few days prior and during its eruption by Earth-based assets such as Solar Dynamics Observatory, IRIS and Hinode.
Its position from the Sun-Earth line was N12W68. The flare recorded by GOES soft-X ray observations indicates a start at 12:56:00, with a peak at 13:55:00 and followed by a long duration event.

This event is particularly interesting for several reasons: first, the large coverage available with different instruments allows us to follow the pre-flare phase, during which the filament slowly rises and pushes overlying coronal arcades away, as modelled in 3D numerical simulations of eruptive flares. This may be linked to the observed large-scale reconfiguration reported in section~\ref{sec:reconfig}. Second, Doppler velocity and intensity changes in several lines are reported between the upper chromosphere and transition regions (SPICE diagnostics) as well as coronal lines (Hinode/EIS diagnostics) with different view points. This is the first time that such an event is seen stereoscopically with different spectrometers.  Finally, the extended coverage, from spectroscopy to EUV and X-Ray imaging allows us to understand the evolution of the magnetic field changes during the different phases of the flare. A dedicated study of this event is available in Janvier et al. 2022 (this volume, in prep) and in Ho et al. 2022 (this volume, in prep).

\section{Instrument Performance at perihelion\label{sec:performance}}
\label{section:Performance}

 The pre-flight instrument characterisation  is discussed in telescope specific papers in this issue (Auch\`{e}re and EUI consortium partners 2022 (this volume, in prep);
 Aznar Cuadrado and EUI consortium partners 2022 (this volume, in prep); Gissot and EUI consortium partners 2022 (this volume, in prep)). The period around the 2022 March 27 perihelion was the first time the instrument was operated in the environment for which it was primarily designed. In this section we review how the instrument was operated technically and the resulting performance.
Overall, FSI and \hrieuv performed largely nominally while \hrilya suffered from a temporary degradation in throughput and resolution (see below).

\subsection{Sensors}

The three EUI telescopes share the same CMOS sensor design \citep{Rochus2020}. The sensors consist of two parts of each 1536~x~3072 pixels, stitched together as a 3072~x~3072 array. The HRI sensors are used sub-fielded to 2048~x~2048 pixels. Careful inspection reveals that the stitching line remains visible in all three telescopes but most noticeably in \hrilya
(see Fig.\,\ref{hrilya_fig1} c) at x=180\,Mm).

Each pixel has a high-gain and low-gain read-out which can be brought to the ground independently or selected per pixel per intensity threshold. Onboard electronics then re-scales the high-gain and low-gain signals from different pixels into one coherent intensity range over all pixels in a 'recombined' image.  In the 2022 March/April period,  FSI and \hrieuv have been nominally operated in the combined gain mode, resulting in 15-bit images (an intensity range in Level 1 files of 0--32767 DN for FSI and 0--25600 DN for \hrieuv).  The selection threshold  between low-gain and high-gain read-out happens near a Level 1 intensity level of 1097\,DN for FSI and 1118\,DN for \hrieuv.  This transition is weakly visible in the FSI and \hrieuv images as a band of enhanced noise, which is to be expected given the different photon statistics in the low-gain and high-gain read-out channels.
In contrast, \hrilya has been operated exclusively in low-gain read-out resulting in 12-bit images (an intensity range of 0--4095 DN in the Level 1 image files) and which do not show such transition.

Other sensor artefacts, affecting FSI, are dark vertical bands in very faint areas, aligned with the brightest on-disk features. This effect is assumed to be caused by saturation of the high-gain read-out of pixels in the same column  and is still under investigation. A post-processing semi-empirical fix is being developed.

\subsection{Onboard Processing}

EUI is equipped with software controlled onboard calibration electronics to correct the images pixel-wise for offset and flat field before compression. For FSI, pre-flight offset and flat field maps are available on board and have been applied until 2022 March 16 when it was discovered that the flat field map was not applied correctly. Correction for the flat field was turned off at this time and subsequent FSI images  have only the offset map applied.

For \hrieuv, only a synthetic 4-column pattern is subtracted that mimics the observed offset. For \hrilya no onboard correction is applied.

Despite the close solar proximity, radiation hits on the sensors have been very limited and the onboard cosmic ray corrector has therefore not been employed. Enhanced radiation hits were observed, e.g. following the 2022 March 21 event (see subsection \ref{section:EPDevent}).

\subsection{Image resolution\label{sec:image.resolution}}

\begin{figure}[ht]
\centering
\includegraphics[width=0.48\textwidth]{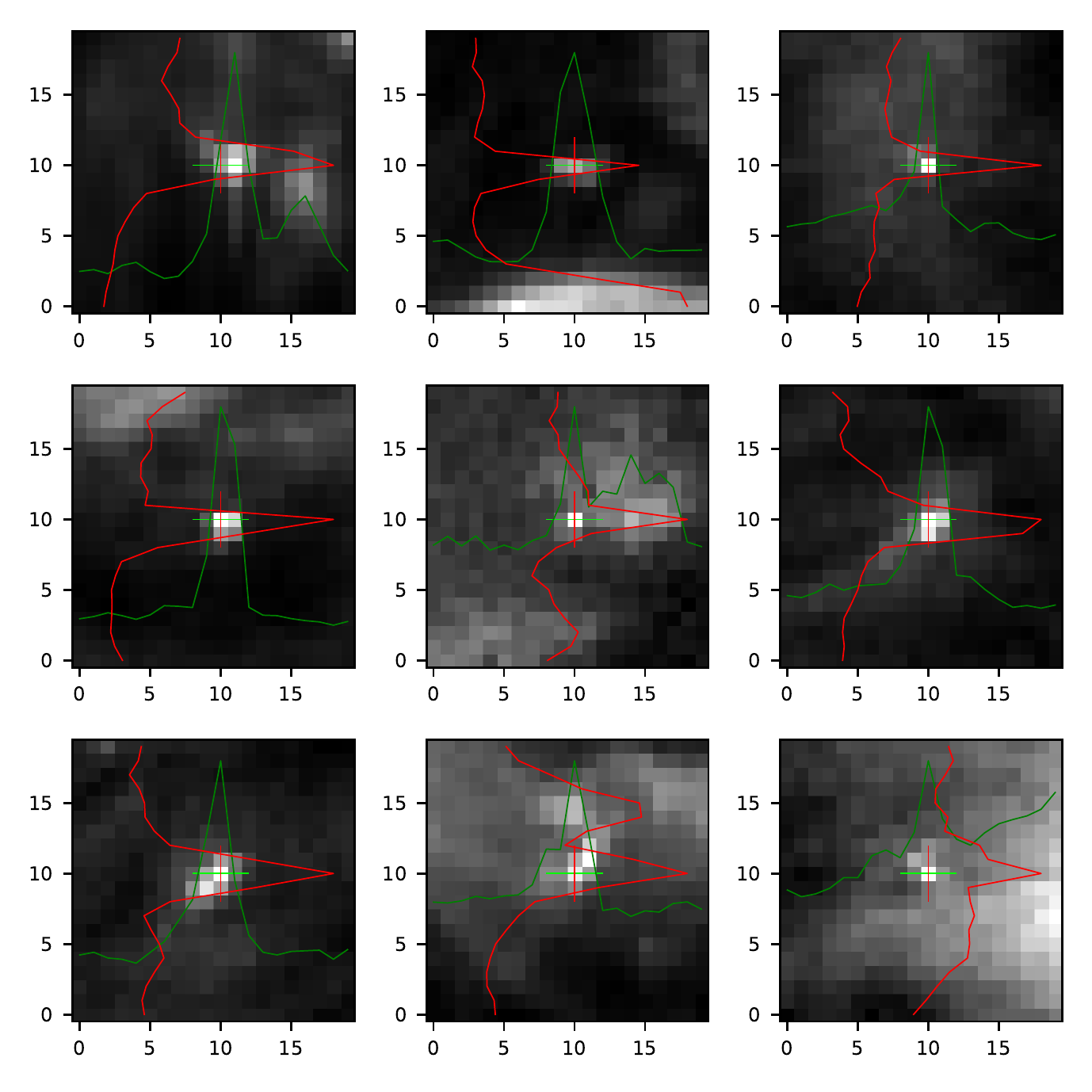}
\caption{Enlargements of selected compact features observed by FSI at 17.4\,nm on 2022 March 7 at 06:20:30 UT. Assuming that the sources are unresolved, the green and red   (horizontal and vertical respectively) profiles indicate a FWHM width of the effective Point Spread Function of 1.5 pixels. See Sect. \ref{sec:image.resolution}.}
\label{fig:fsi_resolution}
\end{figure}

During commissioning, the FWHM of the FSI Point Spread Function (PSF) was estimated to be ~1.5 pixels, or 6.66". As shown by Fig.\,\ref{fig:fsi_resolution}, there is no sign of changes so far. At closest approach, i.e. at 0.3\,au, this corresponds to a resolution of 2.5" (two pixels) as seen from 1\,au, similar to that of STEREO/EUVI \citep[2.4",][]{Wuelser2007}.

\begin{figure}[ht]
\centering
\includegraphics[width=0.48\textwidth]{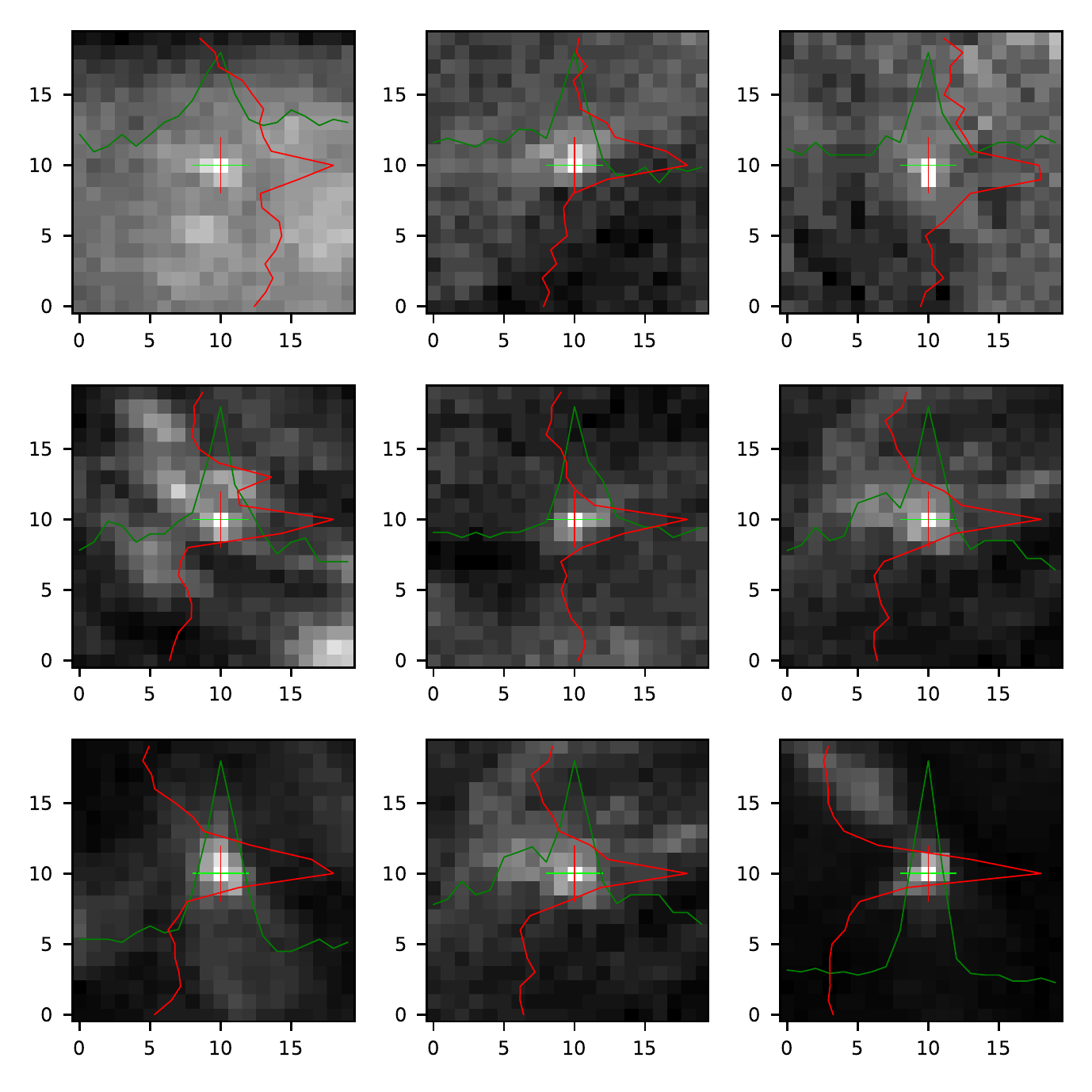}
\caption{Enlargements of selected compact features observed by \hrieuv on 2022 March 7 at 00:41:55 UT. Assuming that the sources are unresolved, the green and red   (horizontal and vertical respectively) profiles indicate a FWHM width of the effective Point Spread Function of 1.5 pixels.  See Sect. \ref{sec:image.resolution}.}
\label{fig:hri_resolution}
\end{figure}

The excellent resolving quality of \hrieuv was confirmed during perihelion through identification of point-like features (Fig.\,\ref{fig:hri_resolution}), and slender features (Fig.\,\ref{slender_fig}), with a FWHM width of about 1.5 pixels. This is consistent with the spatial resolution of the \hrieuv telescope being equal the Nyquist sampling limit of 2 pixels, $2\times 0.492 \arcsec$. At perihelion just inside 0.3~au this corresponds to about $2\times100$~km on the Sun.

\begin{figure}[ht]
\centering
\includegraphics[width=0.4\textwidth]{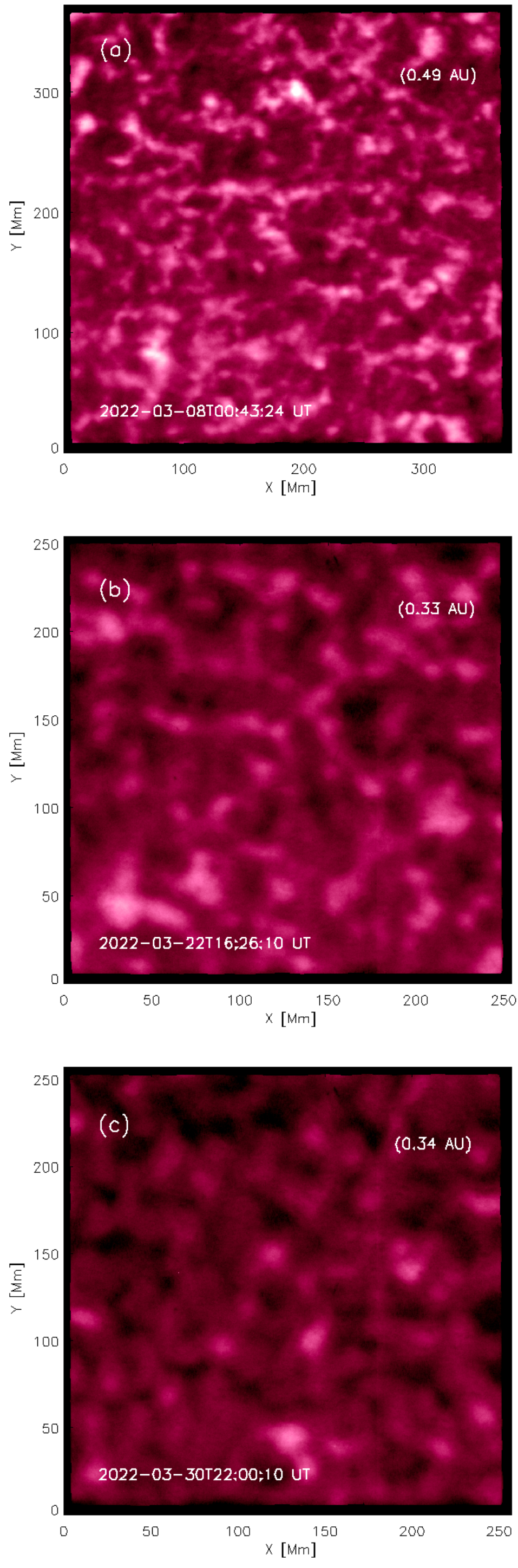}
\caption{\hrilya observations of a set of quiet sun regions located near disk centre, obtained on 2022 March 8 (panel a), 2022 March 22 (panel b), and 2022 March 30 (panel c). See Sect. \ref{sec:image.resolution}.}
\label{hrilya_fig1}
\end{figure}

From the beginning of the mission, the \hrilya spatial resolution  was found to be lower than expected, with a first estimate placing it at around 3\arcsec\, \citep[see][]{2021A&A...656L...4B}. %Berghmans et al. 2021. Campfires paper
However, during the perihelion approach of Solar Orbiter the telescope has shown a further substantial degradation of spatial resolution, contrast and throughput. Fig.\,\ref{hrilya_fig1} shows three images of quiet sun regions taken on 2022 March 8 (where Solar Orbiter was at a distance to the Sun of 0.49\,au), March 22 (at 0.33\,au), and March 30 (at 0.34\,au). All targets were selected to be near disk centre. The loss in performance can be clearly seen in panels (b) and (c),  immediately before and after the closest approach to the Sun on 2022 March 26 (at 0.32\,au).
Most obvious is the resolution degradation which may be a result of a heat effect on the entrance filter of the \hrilya telescope. In addition, as Solar Orbiter approaches the Sun, both the contrast and throughput degrade by approximately 37~\% with respect to data taken before the perihelion (around mid-February 2022), and these recover slowly after perihelion passage.

\subsection{Filters and light leaks}

The reasons for the observed overall loss of performance of \hrilya during perihelion passage are currently under investigation.
Experience from ground testing of the entrance filter with heating to 200 $^\circ$C has revealed a non-linear decline of its transmission of up to 40~$\%$ as a function of temperature. Part of the loss of the channel’s throughput may be associated with the temperature dependency of the filter.
Consistent with this, some  throughput and resolution recovery was observed further from the Sun, on 2022 June 12, which was the first \hrilya observations after 2022 April.
A full assessment of the evolution of throughput since launch and of its comparison with the expectations from ground calibration will be the subject of a separate publication.

In contrast to \hrilya, the EUV channels may be affected by light leaks. Two issues are known to affect the FSI filters:
\begin{itemize}
\item A faint light leak, likely caused by a pinhole in the front filter, affects the images from both channels. Due to the specific design of FSI, its visibility depends on the distance to the Sun and pointing of the spacecraft. There is no quantitative correction for this yet. This has no impact for morphological studies, but care must be taken for photo-metric analysis off-disk.
\item A very faint light-leak, invisible in regular images, affects the 30.4\,nm data taken in coronagraph mode. Its origin is still unknown, but only a small number of images are affected.
\end{itemize}

\subsection{Pointing error and jitter}

The pointing information in the World Coordinate System (WCS) keywords of the EUI Level 1 FITS files are based on the as-flown Solar Orbiter spacecraft kernels. As such, these keywords capture most of Solar Orbiter pointing instabilities, but unfortunately not all. Even after correcting for the known pointing variation, occasional jitter remains visible from image to image (as well as slower trends) in high cadence \hrieuv sequences. But in general, \hrieuv images do not seem to be affected much by jitter blurring.

For FSI images, in which the solar limb is always visible, the WCS pointing keywords are updated in the EUI Level 2 FITS files with much more precise information from a procedure that fits a circle to the solar limb. This is unfortunately not possible for \hrieuv and \hrilya image sequences and the data user is advised to use alignment methods to remove the remaining jitter.

\section {Conclusions\label{sec:conclusions}}
\label{section:Conclusions}

During the Solar Orbiter perihelion passage of 2022 March 26, and the weeks before and after, EUI collected more than 35000 images.  Solar Orbiter reached a distance to the Sun as low as 0.32\,au, closer to the Sun than any other coronal imager. Both FSI and \hrieuv operated at design specifications during the perihelion passage but \hrilya suffered from an unexpected (but reversible) performance degradation near perihelion that needs to be studied further.

EUI has achieved the highest resolution images ever of the solar corona in the quiet Sun and polar coronal holes. Ubiquitous EUV brightenings (a.k.a. campfires) and small scale jets were recovered down to the resolution limit of \hrieuv of about 200\,km on the solar surface. These smallest features require further investigations for their relevance to the heating of the corona and the powering of the solar wind.

Whereas the Hi-C sounding rocket \citep{2014SoPh..289.4393K,2019SoPh..294..174R} achieved comparable resolution in active regions, \hrieuv imaged  active regions at much longer sequence durations (hours) at high cadence (3~s). Known phenomena such as coronal braiding, decayless oscillations, coronal rain and flaring activity were observed in unprecedented details.

The highest resolution full disc image ever was constructed as a mosaic of 25 high resolution images. Together with the PHI and SPICE instruments onboard Solar Orbiter, this full disc mosaic will be repeated twice per year when Solar Orbiter crosses a distance of 0.5\,au from the Earth.

Meanwhile, the big novelty of FSI, namely its very extended FOV, allowed the imaging of eruptions off limb further than ever before, with in particular the prominence eruption showing a bewildering variety in structural appearances.

Future perihelia will go another 10~\% closer to the Sun, to a distance of 0.29\,au from the Sun, and as the mission progresses, Solar Orbiter/EUI will also observe from increasing solar latitudes.   Many of the SOOPs and EUI observations presented in this paper will be repeated from these upcoming vantage points. Special attention will be paid to  deepening joint observations with other instruments on Solar Orbiter but also with Earth-bound observatories in space and on the ground.

This paper presented how the EUI observations contributed to the various Solar Orbiter Observations Programs (SOOPs) that implement cross-instrument science goals. By  highlighting particular features and events, many of which require further study, this paper intended
 to demonstrate the potential of the EUI data and to inspire external users to take part in the EUI data analysis.
The EUI dataset presented in this paper has been distributed as part of the EUI Data Release 5.0 \citep{euidatarelease5} and is freely accessible. We encourage EUI data users to read the release notes and get in contact with the EUI team for specific support.

\begin{acknowledgements}
The building of EUI was the work of more than 150 individuals during more than 10 years. We gratefully acknowledge all the efforts that have led to a successfully operating instrument.
The authors thank the Belgian Federal Science Policy Office (BELSPO) for the provision of financial support in the framework of the PRODEX Programme of the European Space Agency (ESA) under contract numbers 4000112292, 4000134088, 4000134474, and 4000136424. The French contribution to the EUI instrument was funded by the French Centre National d'Études Spatiales (CNES); the UK Space Agency (UKSA); the Deutsche Zentrum f\"ur Luft- und Raumfahrt e.V. (DLR); and  the Swiss Space Office (SSO). PA and DML acknowledge funding from STFC Ernest Rutherford Fellowships No. ST/R004285/2 and ST/R003246/1, respectively. SP acknowledges the funding by CNES through the MEDOC data and operations center. L.P.C. gratefully acknowledges funding by the European Union. Views and opinions expressed are however those of the author(s) only and do not necessarily reflect those of the European Union or the European Research Council (grant agreement No 101039844). Neither the European Union nor the granting authority can be held responsible for them.
\end{acknowledgements}

\bibliographystyle{aa}
\bibliography{bibliography}

\begin{appendix}
\label{s-appendixA}
\section{EUI Data set characteristics}

\begin{table*}
\caption{Summary of SOOPs and corresponding EUI datasets. In between the SOOPs, additional FSI synoptic images have been taken that are not listed in this table. Some specific calibration datasets have also been omitted.}\label{datasetTable}
\centering
\begin{tabular}{lrrlcl}
\hline\hline
   & \# images & cadence & start & end  & comment\\
\hline\hline
\multicolumn{3}{l}{\bf L\_SMALL\_MRES\_MCAD\_Connection-Mosaic\phantomsection\label{Connection-Mosaic}} &  2022-03-01 18:00 & 2022-03-03 03:21 & 3 pointings \\
\hrieuv &  810 &  2\,min & 2022-03-02 00:00 & 2022-03-03 03:00  & \\
\hrilya &  809 & 2\,min & 2022-03-02 00:00 & 2022-03-03 03:00 & \\
FSI174 &   108 & 15\,min & 2022-03-02 00:01 & 2022-03-03 02:45 &  \\
FSI304 &   108 & 15\,min & 2022-03-02 00:01 & 2022-03-03 02:46 & \\
\hline
\multicolumn{3}{l}{\bf L\_SMALL\_MRES\_MCAD\_Connection-Mosaic} &  2022-03-30 07:55 & 2022-03-31 17:40 & 6 pointings\\
\hrieuv \& &   576 & 30\,s  & 2022-03-30 11:00 & 2022-03-31 15:47 & 6 bursts of 48\,min at 11:00, 18:00,   \\
\hrilya &   288 & 60\,s  &  &  &    22:00, 03:30, 09:00, 15:00 \\
FSI174 &   188 & 10\,min  & 2022-03-30 08:00 & 2022-03-31 17:30 &  \\
FSI304 &   64 & 30\,min  & 2022-03-30 08:00 & 2022-03-31 17:30 & \\
\hline
\multicolumn{3}{l}{\bf L\_SMALL\_HRES\_HCAD\_Slow-Wind-Connection\phantomsection\label{Slow-Wind-Connection}}       &  2022-03-03 06:00 & 2022-03-06 16:45 & various pointings \\
\hrieuv \& & 3x720 &  5\,s & 2022-03-03 09:40 & 2022-03-05 16:20 & 1h bursts starting  3th 09:40,\\
\hrilya &3x720 &  5\,s &&& 4th 10:45, 5th 15:20 \\
FSI174 &  479 & 10\,min & 2022-03-03 06:00 & 2022-03-06 15:11 &  \\
FSI304 &  160 & 30\,min & 2022-03-03 06:00 & 2022-03-06 15:01 &  \\
\hline
\multicolumn{3}{l}{\bf L\_SMALL\_HRES\_HCAD\_Slow-Wind-Connection}       &  2022-03-17 06:00 & 2022-03-22 00:00 & various pointings \\
\hrieuv \& & 5x720 & 5\,s &2022-03-17 09:47 & 2022-03-21 12:36 & 1h bursts: 17th 09:47, 18th 10:10, \\
\hrilya    & 5x720 & 5\,s &                 &   & 19th 10:36, 20th 11:27, 21th 11:36 \\
FSI174 & 630 & 10\,min &2022-03-17 06:00 & 2022-03-21 23:51 & \\
FSI304 & 210 & 30\,min &2022-03-17 06:00 & 2022-03-21 23:30 & \\
\hline
\multicolumn{3}{l}{\bf R\_SMALL\_HRES\_MCAD\_Polar-Observations\phantomsection\label{Polar-Observations}}         &  2022-03-06 16:45 & 2022-03-06 21:50 & pointing: North pole\\
\hrieuv & 7, 149  & 2\,s, 30\,s &2022-03-06 17:34 & 2022-03-06 18:51 & \\
\hrilya & 105  & 60\,s &2022-03-06 17:05 & 2022-03-06 18:50 & \\
FSI174 & 75   & 90\,s &2022-03-06 17:05 & 2022-03-06 21:21 & \\
FSI304 & 5    & 30\,min & 2022-03-06 19:20 & 2022-03-06 21:20 & \\
\hline
\multicolumn{3}{l}{\bf R\_SMALL\_HRES\_MCAD\_Polar-Observations}         &  2022-03-30 03:30 & 2022-03-30 07:00 &  pointing: South Pole \\
\hrieuv & 600 &  3\,s & 2022-03-30 04:30 & 2022-03-30 05:00 &  \\
\hrilya & 360 & 5\,s & 2022-03-30 04:30 & 2022-03-30 05:00 & \\
FSI174 & 18 & 10\,min &2022-03-30 03:30 & 2022-03-30 07:01 & \\
FSI304 & 7 & 30\,min & 2022-03-30 03:50 & 2022-03-30 07:01 & regular spacing\\
\hline
\multicolumn{3}{l}{\bf R\_SMALL\_HRES\_MCAD\_Polar-Observations}         &  2022-04-04 16:25 & 2022-04-05 23:53 &  pointing: North Pole\\
\hrieuv & 90,755 & 30\,s \&  60\,s & 2022-04-04 16:25 & 2022-04-05 05:44 & variable cadence\\
\hrilya & 786 & 60\,s & 2022-04-04 16:25 & 2022-04-05 05:44 &  \\
FSI174 & 190  & 10\,min &2022-04-04 16:30 & 2022-04-05 23:50 & 5\,min cadence in last 3h\\
FSI304 & 51  & 30\,min &2022-04-04 16:30 & 2022-04-05 20:30 & \\
\hline
\multicolumn{3}{l}{\bf R\_BOTH\_HRES\_HCAD\_Nanoflares\phantomsection\label{Nanoflares}}                  &  2022-03-06 21:50 & 2022-03-07 03:00 & pointing: active region\\
\hrieuv & 7, 1188, 149, 60 & 2, 5, 12, 20\,s &2022-03-07 00:29 & 2022-03-07 03:00 & variable cadence\\
\hrilya & 783, 60 &  12, 20\,s&2022-03-07 00:00 & 2022-03-07 03:00 & variable cadence\\
FSI174 & 354 & 30\,s &2022-03-07 00:00 & 2022-03-07 03:00 & \\
\hline
\multicolumn{3}{l}{\bf R\_BOTH\_HRES\_HCAD\_Nanoflares}                  &  2022-03-08 00:00 & 2022-03-08 03:00 & pointing: disc center, quiet Sun \\
\hrieuv & 588, 1188, 149, 60  & 3, 5, 12, 20\,s &2022-03-08 00:00 & 2022-03-08 03:00 & variable cadence\\
\hrilya & 783, 60 & 12, 20\,s &2022-03-08 00:00 & 2022-03-08 03:00 & variable cadence \\
FSI174 & 355 & 30\,s &2022-03-08 00:00 & 2022-03-08 03:00 & \\
\hline
\multicolumn{3}{l}{\bf R\_BOTH\_HRES\_HCAD\_Nanoflares}                  &  2022-03-17 00:00 & 2022-03-17 02:55 & quiet Sun \\
\hrieuv & 600 & 3\,s &2022-03-17 00:18 & 2022-03-17 00:48 & \\
\hrilya & 150 & 12\,s &2022-03-17 00:18 & 2022-03-17 00:48 & \\
FSI304 & 28 & 60\,s &2022-03-17 00:03 & 2022-03-17 01:04 & gap between 00:16 and 00:50 \\
\hline
\multicolumn{3}{l}{\bf R\_BOTH\_HRES\_HCAD\_Nanoflares}                  &  2022-03-17 03:00 & 2022-03-17 05:55 & pointing: active region \\
\hrieuv & 900 & 3\,s &2022-03-17 03:18 & 2022-03-17 04:03 &\\
\hrilya & 348 & 5\,s &2022-03-17 03:18 & 2022-03-17 03:47 & some images missing\\
FSI304 & 28  & 60\,s &2022-03-17 03:03 & 2022-03-17 04:03 & gap between 03:16 and 03:50\\
\hline
\multicolumn{3}{l}{\bf R\_BOTH\_HRES\_HCAD\_Nanoflares}                  &  2022-03-30 00:00 & 2022-03-30 03:24 & pointing: active region \\
\hrieuv & 900 & 3\,s &2022-03-30 00:03 & 2022-03-30 00:48 & \\
\hrilya & 360 & 5\,s &2022-03-30 00:18 & 2022-03-30 00:48 & \\
FSI174 & 14 & 10\,min &2022-03-30 01:00 & 2022-03-30 03:20 & 03:10 missing\\
FSI304 & 5 & 30\,min &2022-03-30 01:00 & 2022-03-30 03:20 & \\
\hline
\end{tabular}
\end{table*}

\begin{table*}
\caption{Summary of SOOP instances and corresponding EUI datasets. In between the SOOPs, additional FSI synoptic images have been taken that are not listed in this table. Some specific calibration datasets have also been omitted.}\label{datasetTable2}
\centering
\begin{tabular}{lrrlcl}
\hline\hline
   & \# images & cadence & start & end  & comment\\
\hline\hline
\multicolumn{3}{l}{\bf Full Disc Mosaic\phantomsection\label{Full Disc Mosaic}}                                 &  2022-03-07 07:00 & 2022-03-07 11:30 & 25 pointings \\
\hrieuv & 450 & - &2022-03-07 07:01 & 2022-03-07 11:30 & 18 images/pointing, HG/LG\\
\hrilya & 200 & - &2022-03-07 07:04 & 2022-03-w07 11:29 & 8 images/pointing\\
FSI174 & 25 & 11\,min &2022-03-07 07:05 & 2022-03-07 11:30 & \\
FSI304 & 25 & 11\,min &2022-03-07 07:05 & 2022-03-07 11:30 & \\
\hline
\multicolumn{3}{l}{\bf L\_FULL\_HRES\_MCAD\_Coronal-He-Abundance\phantomsection\label{Coronal-He-Abundance}  }       &  2022-03-07 16:00 & 2022-03-07 20:00 & \\
FSI174 & 8  & 30\,min &2022-03-07 16:00 & 2022-03-07 19:31 & HG, occulted, exptime 1000s\\
\hline
\multicolumn{3}{l}{\bf R\_BOTH\_HRES\_MCAD\_Bright-Points} \phantomsection\label{BrightPoints}             &  2022-03-08 08:10 & 2022-03-08 16:45 & pointing: disc center\\
\hrieuv & 120 & 1\,min &2022-03-08 08:10 & 2022-03-08 10:10 & \\
\hrilya & 120 & 1\,min &2022-03-08 08:10 & 2022-03-08 10:10 & \\
FSI174 & 77  & 5\,min &2022-03-08 08:10 & 2022-03-08 16:41 & 5\,min cadence till 14:05\\
FSI304 & 74  & 5\,min &2022-03-08 08:10 & 2022-03-08 16:30 & 5\,min cadence till 14:05 \\
\hline
\multicolumn{3}{l}{\bf L\_FULL\_HRES\_HCAD\_Coronal-Dynamics}
\phantomsection\label{Coronal-Dynamics}           &  2022-03-22 03:10 & 2022-03-22 16:30 & pointing: disc center\\
\hrieuv & 1600 & 30\,s &2022-03-22 03:10 & 2022-03-22 16:30 & \\
\hrilya & 800 & 60\,s &2022-03-22 03:10 & 2022-03-22 16:30 & \\
FSI174 & 68 & 10\,min &2022-03-22 03:18 & 2022-03-22 16:21 & gap 15:00-16:00 \\
FSI304 & 68 & 10\,min &2022-03-22 03:18 & 2022-03-22 16:21 & gap 15:00-16:00\\
\hline
\multicolumn{3}{l}{\bf L\_FULL\_HRES\_HCAD\_Coronal-Dynamics }          &  2022-03-27 19:40 & 2022-03-28 16:30 & pointing: disc center\\
\hrieuv & 1249 & 60\,s &2022-03-27 19:40 & 2022-03-28 16:30 & \\
\hrilya & 795 & 60\,s &2022-03-27 19:40 & 2022-03-28 09:00 & \\
FSI174 & 114 & 10\,min &2022-03-27 19:48 & 2022-03-28 16:26 & gap 15:00-16:00 \\
FSI304 & 114 & 10\,min &2022-03-27 19:48 & 2022-03-28 16:26 & gap 15:00-16:00\\
\hline
\multicolumn{3}{l}{\bf L\_BOTH\_HRES\_LCAD\_CH-Bounday-Expansion}\phantomsection\label{CH-Boundary-Expansion}          &  2022-03-25 19:40 & 2022-03-27 00:00 & pointing: disc center\\
FSI174 \&  & 140 & 10\,min & 2022-03-25 19:40 & 2022-03-27 00:00 & gap 04:00-05:00, ... \\
FSI304    & 140  & 10\,min &               &                     & ..., 15:00-16:00, 16:30 -19:30 \\
\hline
\multicolumn{3}{l}{\bf L\_FULL\_HRES\_HCAD\_Eruption-Watch}\phantomsection\label{Eruption-Watch}          &  2022-03-22 19:40 & 2022-03-23 16:30 & pointing: disc center\\
\hrieuv \&& 60 & 30\,s &2022-03-22 03:30 & 2022-03-23 16:30 & 30\,min burst ...\\
\hrilya & 30 & 60\,s &   &   & ... at 03:30, 16:00 \\
FSI174 & 188 & 6\,min &2022-03-22 19:40 & 2022-03-23 16:25 & gap 15:00-16:00 \\
FSI304 & 188 & 6\,min &2022-03-22 19:40 & 2022-03-23 16:25 & gap 15:00-16:00\\
\hline
\multicolumn{3}{l}{\bf L\_FULL\_HRES\_HCAD\_Eruption-Watch}          &  2022-03-29 03:10 & 2022-03-30 00:00 & pointing: disc center\\
\hrieuv \&& 80 & 30\,s &2022-03-29 12:00 & 2022-03-29 12:40 & 30\,min burst ...\\
\hrilya & 40 & 60\,s &   &   & ... at 03:30, 16:00 \\
FSI174 & 208 & 6\,min &2022-03-29 03:10 & 2022-03-29 23:53 & gap 15:00-16:00 \\
FSI304 & 208 & 6\,min &2022-03-29 03:10 & 2022-03-29 23:53 & gap 15:00-16:00\\
\hline
\multicolumn{3}{l}{\bf R\_SMALL\_MRES\_MCAD\_AR-Long-Term\phantomsection\label{AR-Long-Term}}         &  2022-03-31 17:45 & 2022-04-04 20:26 & pointing: Active Region \\
\hrieuv \& & 4x450& 10\,s &2022-04-01 09:19 & 2022-04-04 10:34 & 75\,min burst at 09:19 on ...\\
\hrilya & 4x150  & 30\,s &  &   & ...April 1, 2, 3, 4\\
FSI174 & 538 & 10\,min &2022-03-31 17:50 & 2022-04-04 20:26 & gap April 4 18:30-20:00\\
FSI304 & 179 & 30\,min &2022-03-31 18:00 & 2022-04-04 20:01 & 30\,min cadence till April 4 18:30 \\
\hline
\end{tabular}
\end{table*}
\end{appendix}

\end{document}